\documentclass[reprint,nofootinbib,showpacs,showkeys,prd]{revtex4-1}
\usepackage{graphicx}
\usepackage{epsfig}
\usepackage{amsmath}
\usepackage{amssymb}
\usepackage{calc}
\begin{document}
\title{Scanning space-time with patterns of entanglement}
\author{P\'eter L\'evay and Bercel Boldis}
\affiliation{MTA-BME Quantum Dynamics and Correlations Research Group, Department of Theoretical Physics,
Budapest University of  Technology and Economics, 1521 Budapest, Hungary}

\date{\today}
\begin{abstract}
In the ${\rm AdS}_3/{\rm CFT}_2$ setup we elucidate how gauge invariant boundary patterns of entanglement of the CFT vacuum are encoded into the bulk via the coefficient dynamics of an $A_{N-3}$, $N\geq 4$ cluster algebra. In the static case this dynamics of encoding manifests itself in kinematic space, which is a copy of de Sitter space ${\rm dS}_2$, in a particularly instructive manner.      
For a choice of partition of the boundary into $N$ regions the patterns of entanglement, associated with conditional mutual informations of overlapping regions, are related to triangulations of geodesic $N$-gons. Such triangulations are then mapped to causal patterns in kinematic space.   
For a fixed $N$ the space of all causal patterns is related to the associahedron ${\mathcal K}^{N-3}$ an object well-known from previous studies on scattering amplitudes.
On this space of causal patterns cluster dynamics acts by a recursion provided by a Zamolodchikov's $Y$-system of type $(A_{N-3},A_1)$. We observe that the space of causal patterns is equipped with a partial order, and is isomorphic to the Tamari lattice. 
The mutation of causal patterns can be encapsulated by a walk of $N-3$ particles interacting in a peculiar manner in the past light cone of a point of ${\rm dS}_2$. 

\end{abstract}

\pacs{04.60.-m, 11.25.Hf, 02.40.-k, 03.65.Ud, 03.65.NK, 03.65.Vf, 03.65.Ta}
\keywords{${\rm AdS}_3/{\rm CFT}_2$ correspondence, Cluster algebras,
   Kinematic Space, Quantum Entanglement}

\maketitle
\section{Introduction}

According to a recent exciting idea entanglement patterns associated with certain boundary states of a Conformal Field Theory (CFT) can manifest
themselves in the classical geometry of the bulk\cite{RT,RT2,HRT,Raam1,Raam2,Raam3}.
This opens up the possibility of identifying space-time structures as ones emerging holographically from entanglement data.

Recently apart
from the basic spaces featuring any holographic consideration, namely the bulk and its boundary, another space
called kinematic space has shown up which is usually regarded as
an intermediary between bulk and boundary\cite{Czech1,Czech1b,Myers,Czech2,Czechmera}. However, apart from translating between the language of
quantum information of the boundary to the language of geometry of the bulk, the kinematic space
is also an interesting space in its own right.
For example in the ${\rm AdS}_3/{\rm CFT}_2$ setup taking the static slice of ${\rm AdS}_3$ kinematic space $\mathbb K$ is just the space of geodesics for this static slice. It turns out that geometrically $\mathbb K$ is a copy of two dimensional de Sitter space ${\rm dS}_2$. Then this picture has given rise to the idea of an emergent  de Sitter from conformal field theory\cite{Myers,Zukowski,Zuk,Zuk2,Callebaut,BoerMyers}.

In our previous paper\cite{Levay} using horocycles and their associated lambda lengths\cite{Penner,Pennerbook} we have given a new geometric meaning to the well-known strong subadditivity relation\cite{NC,CS} for entanglement entropies.
In particular we related two aggregate measures on the boundary detecting how far the infrared degrees of freedom are away from satisfying the strong subadditivity relation to the shear coordinates of geodesic quadrangles in the bulk. This boundary measure, the conditional mutual information\cite{NC,Czech1} encapsulating the monogamy property of entanglement is simply related to the geodesic distance between the geodesics forming the opposite sides of the quadrangle, ones having time-like separation as points in kinematic space.
In the dual picture provided by kinematic space the conditional mutual information is also related to the proper time along a timelike geodesic connecting the relevant two points of a causal diamond\cite{Levay,Zhang}.

This observation has revealed a connection with the theory of Teichm\"uller spaces of marked Riemann surfaces\cite{Penner}. In particular we have shown that the two different boundary measures of strong subadditivity are related to the two different triangulations of a geodesic quadrangle. In this picture the shear coordinates\cite{Pennerbook}, satisfying a reciprocal relation, are just possible local coordinates for the space of deformations of quadrangles, i.e. their Teichm\"uller space.
Generalizing these observations for geodesic $N$-gons with $N\geq 4$ an interesting connection with $A_{N-3}$ cluster algebras\cite{ClusterFZ,Williams,Fomin} emerges.  Here the cluster variables are just the lambda lengths\cite{Penner} directly related to the regularized entropies of the boundary via the Ryu-Takayanagi relation\cite{RT}.

We also emphasized that the basic role the gauge invariant conditional mutual informations play in these elaborations
dates back to the unifying role of the Ptolemy relation for geodesic quadrangles\cite{Pennerbook}.
This identity is the basis of recursion relations underlying transformation formulas for shear coordinates of geodesic polygons.
Indeed in the general case of geodesic $N$-gons these recursion relations are precisely of the form of a Zamolodchikov's $Y$-system of $(A_{N-3},A_1)$ type\cite{Zamo,FrenkelSzenes,Ravanini, Gliozzi}.
Moreover, we have also pointed out\cite{Levay} that since boundary intervals with their associated geodesics of the bulk are organized according to the causal structure of their corresponding points in kinematic space\cite{Czech1}, in this manner cluster algebras are also connected to structures of causality in kinematic space.
Finally we have observed that the cluster dynamics based on flips (or alternatively on quiver mutations) provides a dynamics similar to the ones conjectured for holographic codes\cite{Osborne,Yoshida}.

The aim of the present paper is to elaborate on these observations. In particular we would like to give a detailed account of the correspondence between entanglement patterns of the boundary and their encoded images into the bulk and the kinematic space by cluster algebras revealed in\cite{Levay}.

What we find is that the cluster dynamics provides different causal patterns consisting of causal diamonds in kinematic space which are related by cluster mutation. As is well-known\cite{Czech1} the areas of the causal diamonds with respect to the Crofton form correspond to the conditional mutual informations of the boundary. The novelty here is the result that these area labels for the causal diamonds are related to the coefficient variables of a cluster algebra in a simple manner. Moreover, we will see that the space of causal patterns is inherently related to the structure of the associahedron.
One finds that the associahedra encapsulate the many different ways the kinematic space can be scanned by the periodic cluster dynamics encoded in the recursion relations of the corresponding Zamolodchikov $Y$-system. The simplest instance of this dynamics is the additive rule of conditional mutual informations observed in Ref. \cite{Czech1}.
One can conjecture that more intricate manifestations of this dynamics are probably related to some error correction mechanism similar to the ones conjectured for holographic codes\cite{Osborne}.

In this paper for illustrative purposes we will be content with the simplest case based on the CFT vacuum, giving rise to cluster algebras of type $A_{N-3}$ with $N\geq 4$. More general scenarios are left to be explored in future work.
Our hope is that these illustrations will shed some new light on issues concerning holographic codes\cite{Yoshida} and their dynamics\cite{Osborne}, and also initiate further elaborations for exploring connections with related research in scattering amplitudes connected to the ABHY amlituhedron\cite{Nima1} that has come accross the same geometric structure (the associahedron) in a holographic context.

The organization of this paper is as follows. In Section II. we review the basic background on the static slice of ${\rm AdS}_3$ which is represented by the Poincar\'e disk $\mathbb D$. The space of geodesics of $\mathbb D$ is kinematic space $\mathbb K$ which is a copy of de Sitter space ${\rm dS}_2$.
In Section III. we summarize some of the results from Ref.\cite{Levay}. Geodesic quadrangles, their associated lambda lengths, shears and cross ratios appear here, quantities familiar from the theory of marked Riemann surfaces.
The physical interpretation of these quantities is also given here. Then we relate areas in kinematic space to conditional mutual informations in the spirit of Ref.\cite{Czech1} however, with a novel perspective provided by lambda lengths\cite{Penner}.
In Section IV. we present a detailed study on the algebraic structure of triangulations of geodesic polygons. Such triangulations in $\mathbb D$ are mapped to causal patterns in $\mathbb K$. 
Such causal patterns are made of causal diamonds with their associated plaquette variables $X_{j,k}$ where $j=1,2,\dots N$ and $k=1,2,\dots N-3$ introduced. They are related to conditional mutual informations $I_{j,k}$ via Eq.(\ref{lenyeg}). The plaquette variables satisfy a recursion relation (\ref{iksz}), which after a change of variables takes the form of a Zamolodchikov $Y$ system (\ref{ipszilon}).
The solution of this system is well-known\cite{FrenkelSzenes} and can be reinterpreted in terms of conditional mutual informations in terms of some seed variables.
This result connects boundary entanglement structures to the coefficient dynamics\cite{Williams} of a bulk recursive scanning process of $\mathbb K$ governed by an $A_{N-3}$ cluster algebra. 
In Section V. we show that apart from $X_{j,k}$ with $j+k\equiv 0$ mod $2$ there are dual variables with $j+k\equiv 1$ mod $2$. It turns out that the former set of variables is associated with points ($P$), and the latter with complements of the light cones of such $P$s ($\overline C$) in $dS_2$. Moreover, we demonstrate that an interesting relation (Eq.(\ref{huha})) holds between the $P$ and $\overline C$ descriptions. Indeed, the difference between the corresponding mutual informations is related to the geodesic length (proper time) between the time-like separated points in $dS_2$ defining the relevant causal diamond labelled by $P$.

Cluster dynamics unfolds via operations called flips. In Section VI.
we show that
at the boundary level flips are represented by applying the semi-associative law of right or left shifts of binary bracketings introduced by Tamari\cite{Tamari}. This corresponds to a flip in the space of contexts of boundary regions represented by such binary bracketings. At the bulk level they are represented by flips between the two possible diagonals of a geodesic quadrangle. This also corresponds to a flip in the space of triangulations of geodesic $N$-gons. More importantly at the kinematic space level a flip is represented by an elementary step taken by one of $N-3$ particles executing a random walk on the lattice inside a triangular region provided by the past light cone of a distinguished point.
Here we also show that
for fixed $N$ the space of causal patterns $T_{N-2}$ provides a set of regions covering $dS_2$. The collection of such patterns of cardinality given by the Catalan number $C_{N-2}$ can be identified with the vertices of the associahedron ${\mathcal K}^{N-3}$. The boundaries of the associahedron of different dimensions correspond to different partial triangulations. In particular the collection of facets of cardinality $N(N-3)/2$ can be identified with the collection of Ryu Takayanagi geodesics corresponding to diagonals.
Finally we establish on the space of causal patterns a partial order rendering this space a lattice $(T_{N-2},\leq)$ isomorphic to the Tamari lattice.
Our conclusions and an outlook relating our results to current research is left for Section VII.
\vfill

\section{Geodesics and kinematic space}

${\rm AdS}_3$ can be regarded as the hyperboloid in ${\mathbb R}^{2,2}$
characterized by the equation
\begin{equation}
X^2+Y^2-U^2-V^2=-{\ell}^2_{\rm AdS}
\label{ads3}
\end{equation}
where $\ell_{\rm AdS}$ is the AdS radius.
In the following we fix $\ell_{\rm AdS}=1$.
In our considerations we merely take the static (spacelike) slice of ${\rm AdS}_3$ by imposing $V=0$.
We introduce coordinates for this slice as
\begin{subequations}
\begin{align}
X&=X_1=\sinh\varrho\cos\varphi,\\
Y&=X_2=\sinh\varrho\sin\varphi,\\
U&=X_3=\cosh\varrho,
\label{XYZ}
\end{align}
\end{subequations}
with $X^2+Y^2-U^2=-{\ell}^2_{\rm AdS}$.
Here $\varphi\in[0,2\pi]$ and $\varrho\in(-\infty,\infty)$.
Then the set of points satisfying Eqs.(2) is the upper sheet ${\mathbb H}$ of the double sheeted hyperboloid in ${\mathbb R}^{2,1}$ that can also be written as the coset space
$\mathbb H\simeq SO(2,1)/SO(2)$.

After stereographic projection from the point $(X,Y,U)=(0,0,-1)$ of ${\mathbb H}$ to the Poincar\'e disk ${\mathbb D}$ lying in the plane $U=0$ we obtain the coordinates
\begin{equation}
z=\tanh({\varrho}/2)e^{i{\varphi}}=\frac{X+iY}{1+U}=x+iy\in{\mathbb D}.
\label{ze}
\end{equation}
An alternative set of coordinates can be obtained by transforming to the upper half plane $\mathbb U$ by a Cayley transformation
\begin{equation}
\tau=i\frac{1+z}{1-z}=\frac{i-Y}{U-X}=\xi+i\eta\in{\mathbb U},\qquad \eta>0.
\label{tau}
\end{equation}
In our static considerations we will be referring to the spaces $\mathbb D$ and $\mathbb U$ as the bulk, and the unit circle $S^1\simeq\partial\mathbb D$ and the compactified real line ${\mathbb R}{\mathbb P}^1=\mathbb R\cup\{\infty\}\simeq \partial\mathbb U$ as the boundary. 

%\partial\mathbb H helyett \partial\mathbb U?

The metric $ds^2=dX^2+dY^2-dU^2-dV^2$ on ${\mathbb R}^{2,2}$, and the choice of (2) coordinates for the $V=0$ slice yield the induced metric
\begin{equation}
ds^2_{\mathbb H}=d{\varrho}^2+\sinh^2\varrho d\varphi^2
\label{methip}
\end{equation}
on $\mathbb H$.
Its geodesics are given by the formula
\begin{equation}
\tanh\varrho\cos(\varphi-\theta)=\cos\alpha.
\label{geo}
\end{equation}
They are circular arcs in the bulk starting and ending on the boundary see Figure 1.
\begin{figure}
\centerline{\includegraphics[width=5truecm,clip=]{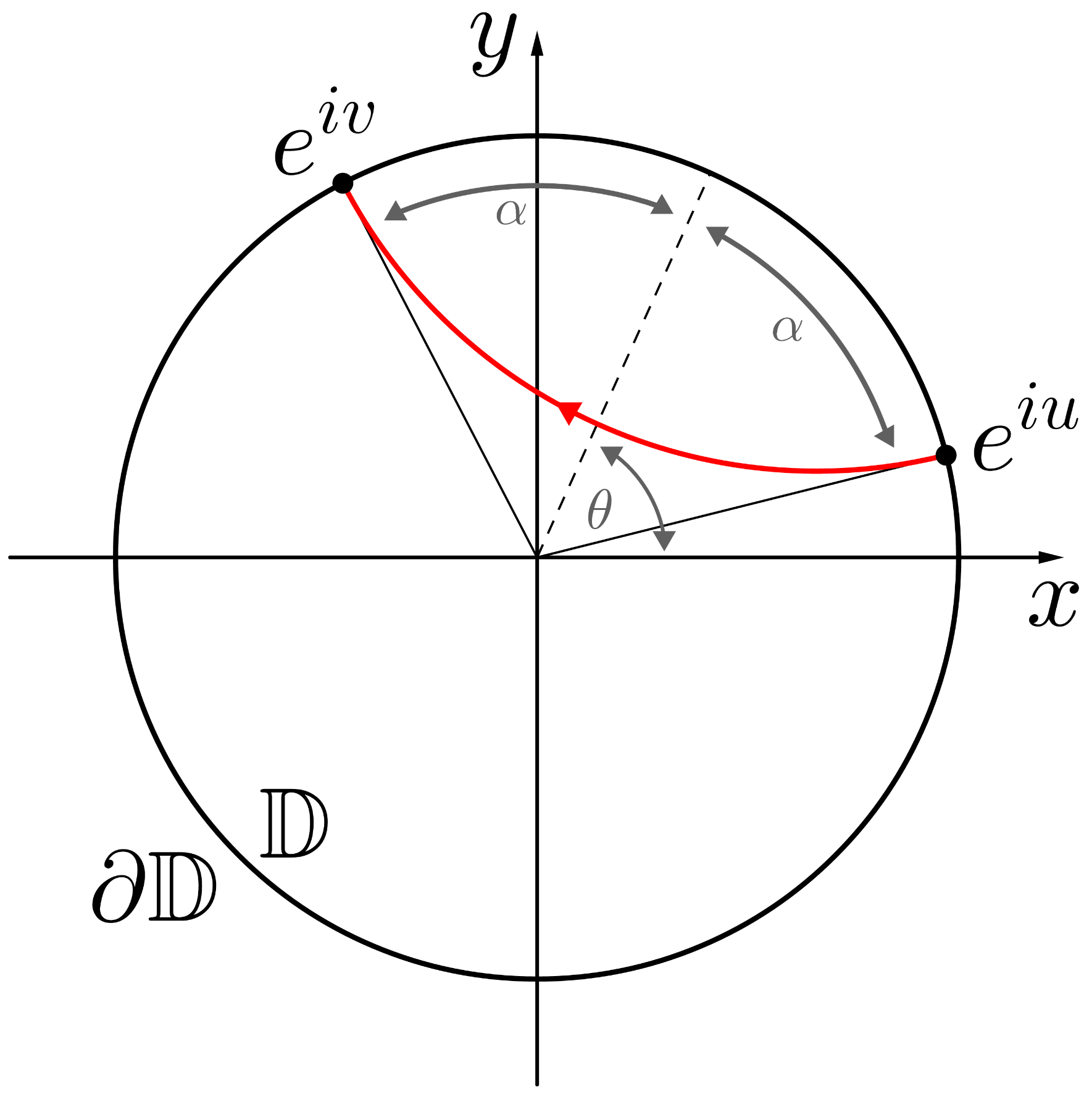}}
\caption{The parametrization of a geodesic. The geodesic (red curve) is parametrized by the pair
$(\theta,\alpha)$ where  $\theta\in[0,2\pi]$ and $\alpha\in [0,\pi]$. The coordinate $\theta$ is the center and $\alpha$ is half the opening angle of the geodesic. Alternatively one can use the pair $(u,v)$ defined by Eq.(\ref{uv}).}
\end{figure}
Here the extra parameters $\theta\in[0,2\pi]$ and $\alpha\in [0,\pi]$ are labelling the geodesics.
Depicted on the disk $\mathbb D$ the coordinate $\theta$ is the center and $\alpha$ is half the opening angle of the geodesic.
Pairs of geodesics differing in orientation are related by $\theta\leftrightarrow \theta+\pi$, $\alpha\leftrightarrow \pi-\alpha$.
Hence our space of geodesics is labelled by the coordinates $(\alpha, \theta)$. It is called the kinematic space\cite{Czech1}.
Topologically the kinematic space $\mathbb K$ is the single sheeted hyperboloid $SO(2,1)/SO(1,1)$ which is the de Sitter space ${\rm dS}_2$.

Note that one can alternatively use the coordinates $(u,v)$ related to the pair $(\alpha,\theta)$
as
\begin{equation}
u=\theta-\alpha,\qquad v=\theta +\alpha.
\label{uv}
\end{equation}
The points $e^{iu},e^{iv}\in\partial{\mathbb D}$ can then be regarded as the starting and the endpoints of a geodesic.
If $\alpha$ is regarded as a time-like coordinate and $\theta$ as a space-like one, then the pair of coordinates $(u,v)$ can be regarded as light cone coordinates.

Since kinematic space described by the coordinates $(\alpha,\theta)$ is a de Sitter space ${\rm dS}_2$ for a geodesic ${\bf A}\in{\mathbb D}$ anchored to a boundary interval $A\in \partial{\mathbb D}$ one can associate a point ${\mathcal A}\in{\mathbb K}$ with three coordinates 
\begin{subequations}\label{haha}
\begin{align}
{\mathcal A}_1&=\frac{\cos\theta}{\sin\alpha}=\cosh\gamma\cos\theta\\\qquad {\mathcal A}_2&=\frac{\sin\theta}{\sin\alpha}=\cosh\gamma\sin\theta\\ \qquad {\mathcal A}_3&=\cot\alpha=-\sinh\gamma
\end{align}
\end{subequations}
where the relation ${\mathcal A}_1^2+{\mathcal A}_2^2-{\mathcal A}_3^2=1$ holds, and we introduced the coordinate transformation
$\cosh\gamma=1/{\sin\alpha}$.
Notice that $\mathbb K$ can be embedded in the same space ${\mathbb R}^{2,1}$ where $\mathbb H$ lives.
Then by employing the coordinates of Eqs.(8) the induced metric this time is of the form
\begin{equation}
ds^2_{\mathbb K}=\frac{d\theta^2-d\alpha^2}{\sin^2\alpha}=\frac{dudv}{\sin^2{\frac{v-u}{2}}}.
\label{kinmetric}
\end{equation}

Notice that by virtue of Eq.(\ref{geo}) our geodesics can alternatively be described by the constraint
${X}\cdot{\mathcal A}=0$, i.e. the vectors ${X}$ and ${\mathcal A }$ are Minkowski orthogonal, with the former is a time-like and the latter is a space-like unit vector.
Then the geodesics are described by the intersection of $\mathbb H$ with a plane in ${\mathbb R}^{2,1}$ with normal vector $\mathcal A$. The two different possible normal vectors give rise to the two points with coordinates $(\theta,\alpha)$ and $(\theta+\pi,\pi-\alpha)$ lying on $\mathbb K$, corresponding to the same geodesic curve with opposite orientations. 
As has been demonstrated in Ref.\cite{BV} the vector ${\mathcal A}$ can be also be regarded as the vector of conserved quantities for the geodesic motion on the pseudosphere
${\mathbb H}$.

 Recall that according to Ref.\cite{Czech1} boundary intervals are organized according to causal structures formed by points located in kinematic space.
 These structures date back to the natural causal structure bulk geodesics enjoy based on the containment relation of boundary intervals. 
If two points in kinematic space are time-like separated then their corresponding geodesics contain each other have no intersection and have the same orientation, and the corresponding boundary intervals are embedded.
If they are null separated then their geodesics have a common endpoint.
Finally if they are spacelike separated, then their corresponding geodesics either have intersection or have different orientation without intersection\cite{Czech1,Zhang}.

\section{Geodesic quadrangles}

Let us fix four points $a,b,c,d$ on the boundary $\partial\mathbb D$ arranged in a counter-clock-wise sense (see Figure 2.). These points give rise to four subregions: $A,B,C,D$ where $A=[a,b], B=[b,c], C=[c,d], D=[d,a]$. Let us also consider the overlapping regions
$E=[a,c]=A\cup B$ and $F=[b,d]=B\cup C$.
Let us denote the oriented geodesics of the bulk $\mathbb D$ anchored to the corresponding boundary regions by ${\bf A,B,C,D,E,F}$. Consider also the corresponding points in kinematic space $\mathbb K$ denoted by ${\mathcal A},{\mathcal B},{\mathcal C},{\mathcal D},{\mathcal E},{\mathcal F}$.
The complement of a boundary region $A$ will be denoted by $\overline{A}$, both of these regions have the same geodesic arc anchored to them however, $\bf A$ and $\overline{\bf A}$ are having different orientation counter clockwise for $\bf A$ and clockwise for $\overline{\bf A}$.
The symbol $\mathcal A$ refers to the corresponding three-vector in ${\mathbb R}^{2,1}$ subject to the constraint ${\mathcal A}_1^2+{\mathcal A}_2^2-{\mathcal A}_3^2 =1$ and $\overline{\mathcal A}$ refers to its negative. 

\begin{figure}
{\includegraphics[height=5truecm]{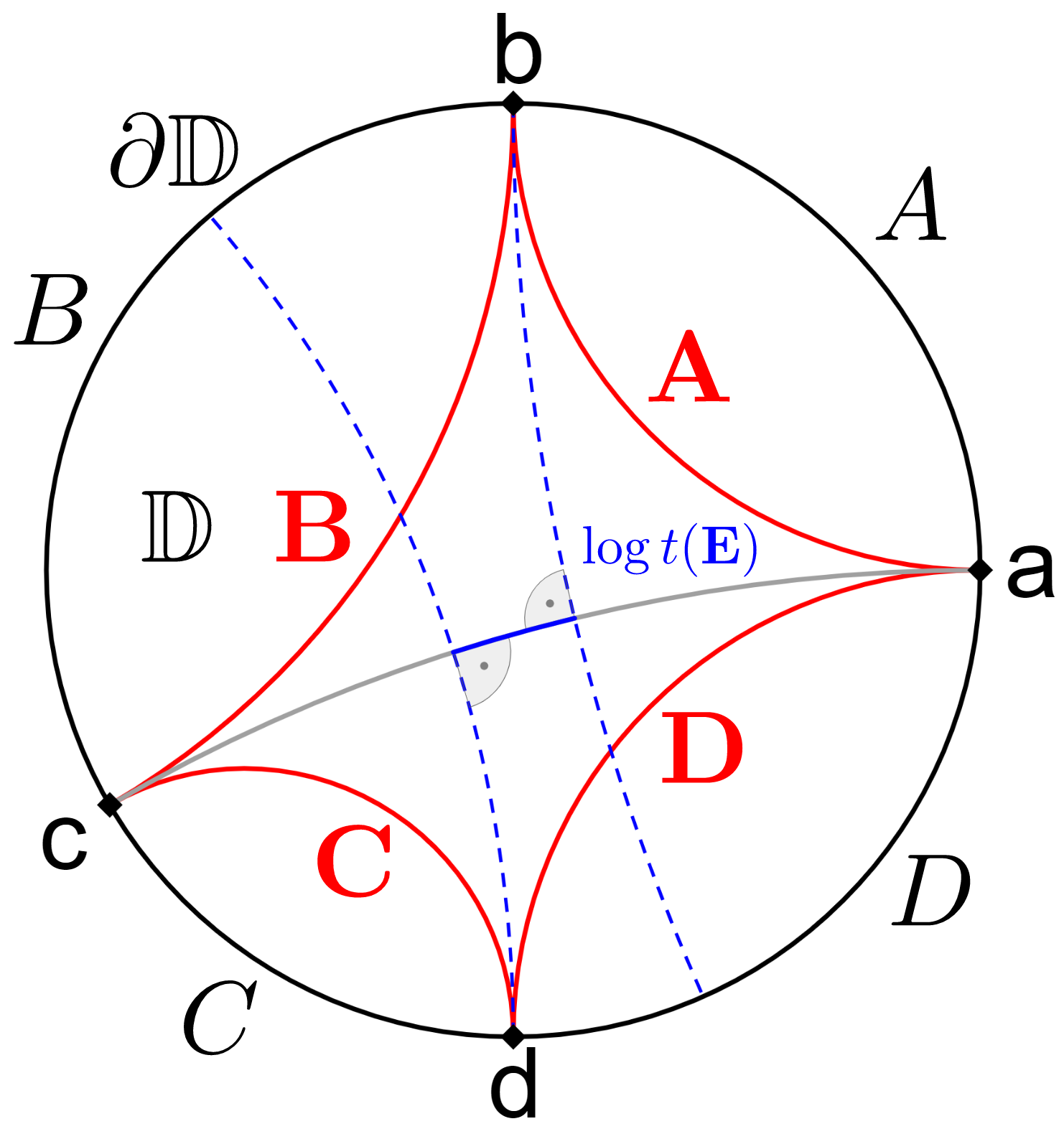}}
\caption{A geodesic quadrangle, with the meaning of the shear coordinate explained.}
\end{figure}

According to strong subadditivity\cite{NC} for subregions $E$ and $F$ of the boundary for the von-Neumann entanglement entropies one has
\begin{equation}
S(E)+S(F)\geq S({E}\cup{{F}})+S({E}\cap{F}).
\label{SSA}
\end{equation}
Note that $\overline{D}=A\cup B\cup C=E\cup F, E\cap F=B$.
Denoting the union of boundary regions by juxtaposition one can also write
\begin{equation}
S(AB)+S(BC)\geq S(B)+S(ABC).
\label{SSA1}
\end{equation}
For pure states one has $S(\overline{D})=S(D)$ and $S(BC)=S(AD)$ hence
\begin{equation}
S(AB)+S(AD)\geq S(B)+S(D).
\label{SSA2}
\end{equation}
Notice that unlike in the classical (Shannon) case in the quantum case it is possible to have either $S(B)>S(AB)$ or $S(D)>S(AD)$. 
In spite of this the sum of these two inequalities can conspire in a manner such that Eq.(\ref{SSA2}) still holds.

One can also express the measures of strong subadditivity in terms of the conditional mutual informations as
\begin{equation}
I(A,C\vert B)=S(AB)+S(BC)-S(B)-S(ABC)\geq 0
\label{Nimacje}
\end{equation}
where 
\begin{equation}
I(A, C\vert B)\equiv S(A \vert B)-S(A\vert BC)\geq 0
\end{equation}
with $S(A\vert B)=S(AB)-S(B)$ is the conditional entropy.
The quantities on the left hand sides indicate that conditioning on a larger subsystem can only reduce the uncertainty about a system.
Alternatively, Eq.(\ref{SSA2}) can be rephrased as
\begin{equation*}
S(A\vert B)+S(A\vert D)\geq 0
\end{equation*}
meaning that subsystem $A$ can be entangled with $B$ reducing $S(AB)$, or with $D$ reducing $S(AD)$, but not both.
This fact is related to the notion of the monogamy of entanglement.

\begin{figure}
\centerline{\includegraphics{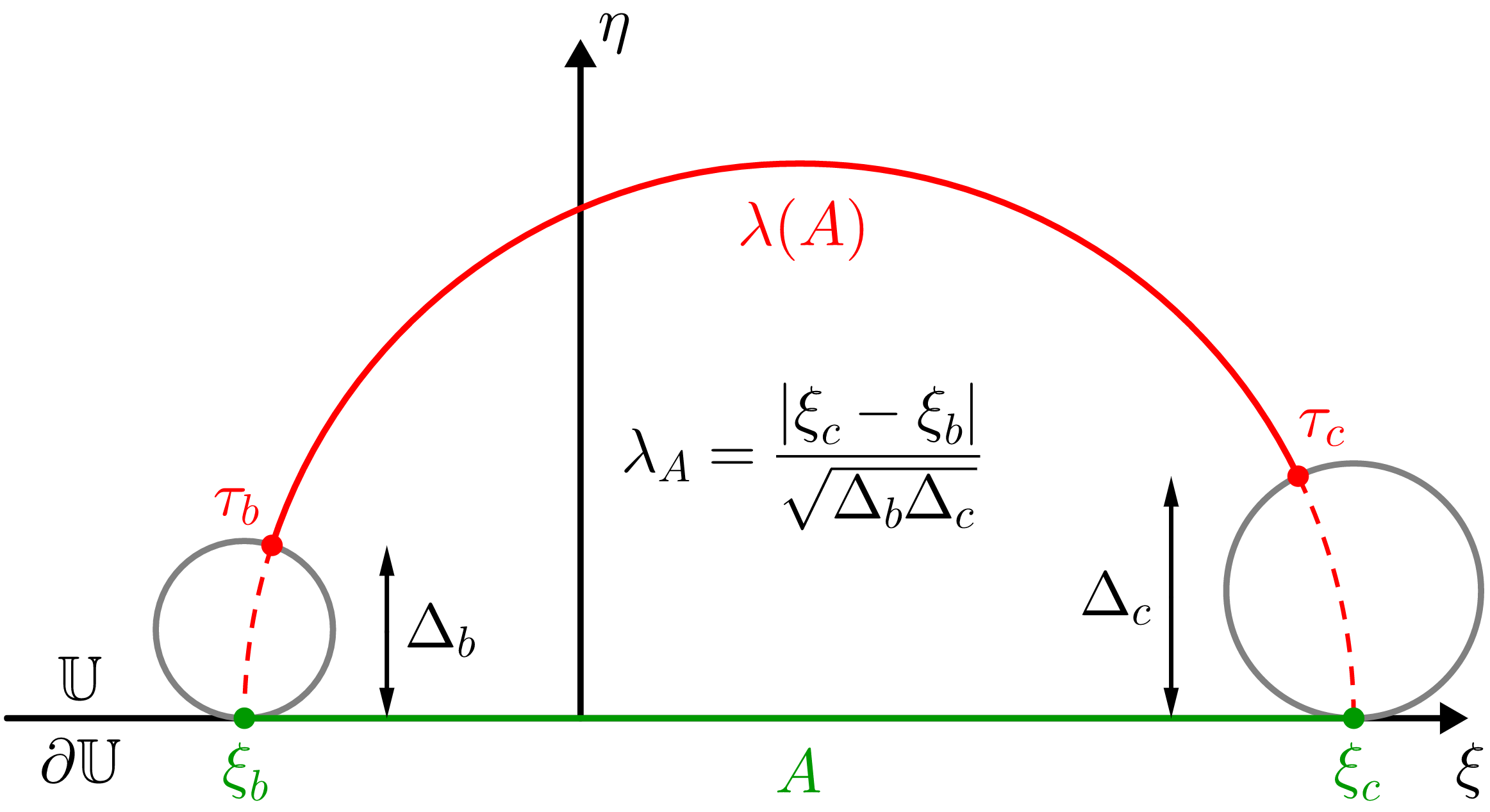}}
\caption{Illustration of the meaning of the lambda length for a circular arc centered on ${\mathbb R}$ which is a part of the boundary $\partial{\mathbb U}$. According to Penner\cite{Penner} the lambda length is the regularized length of the geodesic, where the regulators are horocycles. On $\mathbb U$ these are circles with Euclidean diameters $\Delta$ providing a natural cutoff.}
\end{figure}

In this paper we would like to obtain new insight into the issue of how this boundary data of quantum information is encoded into bulk data of classical geodesic geometry. 
As the first step in achieving this task we recall that, data on conditional mutual information can be expressed in terms of cross ratios\cite{CS,Czech1}.
In order to see this and also to give to this connection a new meaning needed later we proceed as follows.

First we notice that in terms of either coordinates on $\partial{\mathbb U}$ or on $\partial{\mathbb D}$ we have
\begin{equation*}
[d,b;a,c]\equiv
\frac{(\xi_d-\xi_a)(\xi_b-\xi_c)}{(\xi_d-\xi_c)(\xi_b-\xi_a)}=
\frac{\sin\left(\frac{\varphi_d-\varphi_a}{2}\right)
\sin\left(\frac{\varphi_b-\varphi_c}{2}\right)
}
{\sin\left(\frac{\varphi_d-\varphi_c}{2}\right)
\sin\left(\frac{\varphi_b-\varphi_a}{2}\right)}
\label{crossratio}
\end{equation*}
where $[d,b;a,c]$ is the cross ratio
on ${\mathbb R}{\mathbb P}^1$.
Let us then denote the negative of this cross ratio by $t(\bf E)$, and for a boundary subregion $A$ introduce the lambda length\cite{Penner} $\lambda(\bf A)$ of the corresponding geodesic ${\bf A}$. Then in terms of these data one gets the alternative expression\cite{Levay}
\begin{equation}
t({\bf E})=-[d,b;a,c]=\frac{\lambda({\bf B})\lambda({\bf D})}{\lambda({\bf A})\lambda({\bf C})}.
\label{te}
\end{equation}
Note that the lambda length (see Figure 3.) is related to the regularized geodesic length introduced in Ref.\cite{Maxfield}, hence via the Ryu-Takayanagi proposal\cite{RT} to the entanglement entropy of the corresponding boundary interval.
For a detailed discussion on this relationship, see Appendix A. of Ref.\cite{Levay}.

The calculation of entanglement entropy for boundary regions is cutoff dependent. However, the boundary description of the "space of cutoffs" can be transcribed to the bulk description to the "space of horocycles" which is a homogeneous space $\mathbb G$. In Ref.\cite{Levay} we have argued that it is worth regarding the choice of horocycles as a gauge fixing procedure of the gauge degree of freedom noticed in Ref.\cite{Czech2}. In this picture the geometrization of the cutoff dependence of the entanglement entropy corresponds to the horocycle dependence of the lambda length.
In the light of this physical quantities independent of the choice of horocycles (cutoffs) will be called as gauge invariant ones.

Now according to these results $t(\bf E)$ has the dual interpretation as the ratio of Euclidean lengths $\frac{BD}{AC}$ in the boundary $\partial{\mathbb U}$ of the upper half plane, and also as the gauge invariant ratio
of lambda lengths $\frac{\lambda(\bf B)\lambda(\bf D)}{\lambda(\bf A)\lambda(\bf C)}$ in the bulk $\mathbb U$.
More importantly $\log t(\bf E)$ can also be regarded as a gauge invariant shear coordinate\cite{Levay} characterizing the deformation of geodesic 
quadrangles. Alternatively one can regard $\log t(\bf E)$ as a coordinate for the Teichm\"uller space of $\mathbb D$ with four marked points\cite{Penner,Pennerbook}.
For an illustration of the meaning of the shear coordinate see Figure 2.

The geodesics $\bf E$ and $\bf F$ are the diagonals of the geodesic quadrangle $\bf ABCD$. 
One can define a shear coordinate for both of such diagonals. It is easy to show that
\begin{equation}
t({\bf F})=-[b,d;a,c]=\frac{\lambda({\bf A})\lambda({\bf C})}{\lambda({\bf B})\lambda({\bf D})}
\label{Teichm}
\end{equation}
This means that $t({\bf E})t({\bf F})=1$ and these shears serve as alternative local coordinates for the Teichm\"uller 
space of bulk geodesic quadrangles.

The area form associated to the metric of Eq.(\ref{kinmetric}) is related to the Crofton form $\omega$ on kinematic space.
More precisely one has\cite{Czech1}
\begin{equation}
\omega=\frac{\partial^2S(u,v)}{\partial u\partial v}du\wedge dv=\frac{\mathfrak{c}}{12}\frac{du\wedge dv}{\sin^2\left(\frac{v-u}{2}\right)}
\label{Crofton}
\end{equation}
where
\begin{equation}
S(u,v)=\frac{\mathfrak{c}}{3}\log\left(e^{\Lambda}\sin\left(\frac{v-u}{2}\right)\right)
\label{Suv}
\end{equation}
with $e^{\Lambda}$ is the cutoff factor.
Here $\mathfrak{c}$ is the central charge of the boundary CFT related to the bulk Newton constant $G_{\rm N}$ and the AdS length scale $\ell_{\rm AdS}$ via the Brown-Henneaux\cite{BH} relation: $\mathfrak{c}=\frac{3\ell_{\rm AdS}}{2G_{\rm N}}$.

Calculating the integral
\begin{equation}
\int_{\varphi_a}^{\varphi_b}\int_{\varphi_c}^{\varphi_d}\frac{du\wedge dv}{4\sin^2\left(\frac{v-u}{2}\right)}
=\log\left(1+t(\bf F)\right)
\label{integral}
\end{equation}
%csukó zárójel
one proves that the areas calculated with the Crofton form are just the conditional mutual informations, moreover
in terms of these shear coordinates one has\cite{Levay}
\begin{subequations}\label{ssshear5}
\begin{align}
I(A,C\vert B)=\frac{\mathfrak{c}}{3}\log\left(1+t({\bf F})\right)\geq 0\\
I(B,D\vert A)=\frac{\mathfrak{c}}{3}\log\left(1+t({\bf E})\right)\geq 0
\end{align}
\end{subequations}
As a consequence of $t({\bf E})t({\bf F})=1$ one also has
\begin{equation}
I(A,C\vert B)-I(B,D\vert A)=\frac{\mathfrak{c}}{3}\log t({\bf F}).
\label{shearRT}
\end{equation}
This equation gives the kinematic space interpretation of the shear $\log t({\bf F})$ in terms of the difference in areas of the causal diamonds encoding the corresponding conditional mutual informations.

A further geometric interpretation to conditional mutual informations can be given by noticing that\cite{Pennerbook}
 \begin{equation}
 1+t(F)=\cosh^2\frac{\ell}{2},\qquad
 1+t(E)=\cosh^2\frac{\ell^{\prime}}{2}
 \label{otherdist}
 \end{equation}
 where $\ell$ ($\ell^{\prime}$) is the infimum of the geodesic distances between the geodesics ${\bf B}$ and ${\bf D}$ (${\bf A}$ and ${\bf C}$).
Let us then denote by $(\overline{\mathcal A},\dots,\overline{\mathcal F})$ the negatives of the corresponding vectors representing oppositely oriented geodesics in $\mathbb K$.
Then using (\ref{otherdist}) we have
\begin{equation}
\cosh \ell=\overline{\mathcal{B}}\cdot\mathcal{D},\qquad
\cosh \ell^{\prime}=\overline{\mathcal{A}}\cdot\mathcal{C}
\label{kincausal}
\end{equation}
expressing the infimum of the geodesic distances between the pairs of geodesics $(\overline{\bf B},{\bf D})$ and $(\overline{\bf A},{\bf C})$ in $\mathbb D$ in terms of the data of the corresponding pair of points $(\overline{\mathcal{B}},\mathcal{D})$ and
$(\overline{\mathcal{A}},\mathcal{C})$ in $\mathbb K$.
Notice that using the metric of Eq.(\ref{kinmetric}) in $\mathbb K$ one can show\cite{Zhang} that the proper time $\Delta\tau$ between the two points  ${\mathcal D}$ and $\overline{\mathcal B}$ along a timelike geodesic is just the geodesic length $\ell$ between the geodesics ${\bf D}$ and $\overline{\bf B}$ of $\mathbb D$, i.e. we have $\Delta\tau=\ell$.
Likewise the proper time $\Delta\tau^{\prime}$ between the two points ${\mathcal C}$ and $\overline{\mathcal A}$ along a timelike geodesic is just the geodesic length $\ell^{\prime}$ between the geodesics ${\bf C}$ and $\overline{\bf A}$ of $\mathbb D$, i.e. we have $\Delta\tau^{\prime}=\ell^{\prime}$.
Combining these results with the previous ones we obtain the formulae
\begin{subequations}\label{ssshear6}
\begin{align}
I(A,C\vert B)=\frac{\ell_{\rm AdS}}{G_{\rm N}}\cdot\log\cosh\frac{\ell}{2}\\
I(B,D\vert A)=\frac{\ell_{\rm AdS}}{G_{\rm N}}\cdot\log\cosh\frac{{\ell}^{\prime}}{2}
\end{align}
\end{subequations}
These results show that gauge invariant boundary measures of entanglement can be expressed in terms of geometric data of either the bulk ($\ell$) or kinematic space ($\Delta\tau$), see Figure 4.

\begin{figure}
{\includegraphics[width=\columnwidth]{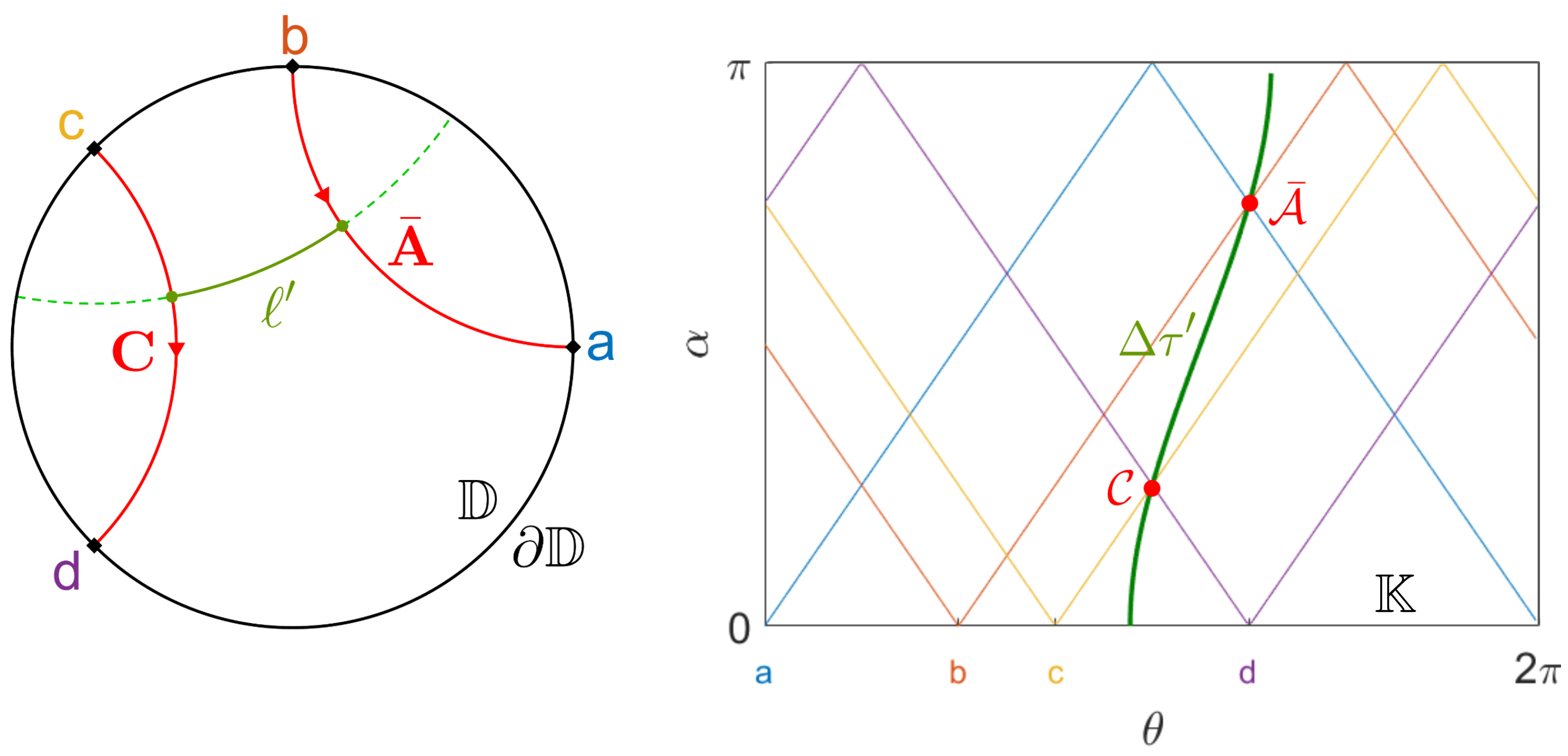}}
\caption{Geodesic length between geodesics in $\mathbb D$
corresponds to proper time in $\mathbb K$ between timelike separated points. Note that according to Eq.(\ref{geo}) the coloured points on $\partial{\mathbb D}$ correspond to coloured point curves (light rays) in $\mathbb{K}$. }
\end{figure}

\section{The algebraic structure of geodesic polygons in kinematic space}

For geodesic $N$-gons we label the $N$ points on the boundary $\partial{\mathbb D}$ by the numbers $\{0,1,2,\dots N-1\}$. By the use of $n=N-3$ diagonals one can obtain a triangulation for these $N$-gons.
The $N$ boundary points give rise to point curves\cite{Czech1} in $\mathbb K$. They are just trajectories of light rays. The set of $N$ left moving and $N$ right moving light rays forms a grid for $\mathbb K$. Our basic idea is to study the algebra of the area labels of causal diamonds, i.e. of the rectangular regions bounded by four different point curves.
We know that the areas of such regions encode conditional mutual information.
We would like to obtain an algebraic characterization of these labelled regions encoding  entanglement patterns of the CFT vacuum.

Similar to the one of Figure 4. now in Figure 5. we see a grid of point curves for $N=4$. However, now we switched to the new labelling $\{a,b,c,d\}\leftrightarrow\{0,1,2,3\}$.
A quadruplet of boundary points forms a quadrangle in the bulk. Choosing a particular diagonal from the two possible ones of this quadrangle gives rise to a shear coordinate. For example for the quadrangle labelled by the set $\{0,1,2,3\}$ we define $u\equiv t_{12,30}=t_{30,12}$ as
\begin{equation}
t_{12,30}=-[1,3;2,0]=
-\frac{\sin\left(\frac{\varphi_1-\varphi_2}{2}\right)
\sin\left(\frac{\varphi_3-\varphi_0}{2}\right)
}
{\sin\left(\frac{\varphi_1-\varphi_0}{2}\right)
\sin\left(\frac{\varphi_3-\varphi_2}{2}\right)}
\label{teshear}
\end{equation}

\begin{figure}
{\includegraphics[width=\columnwidth]{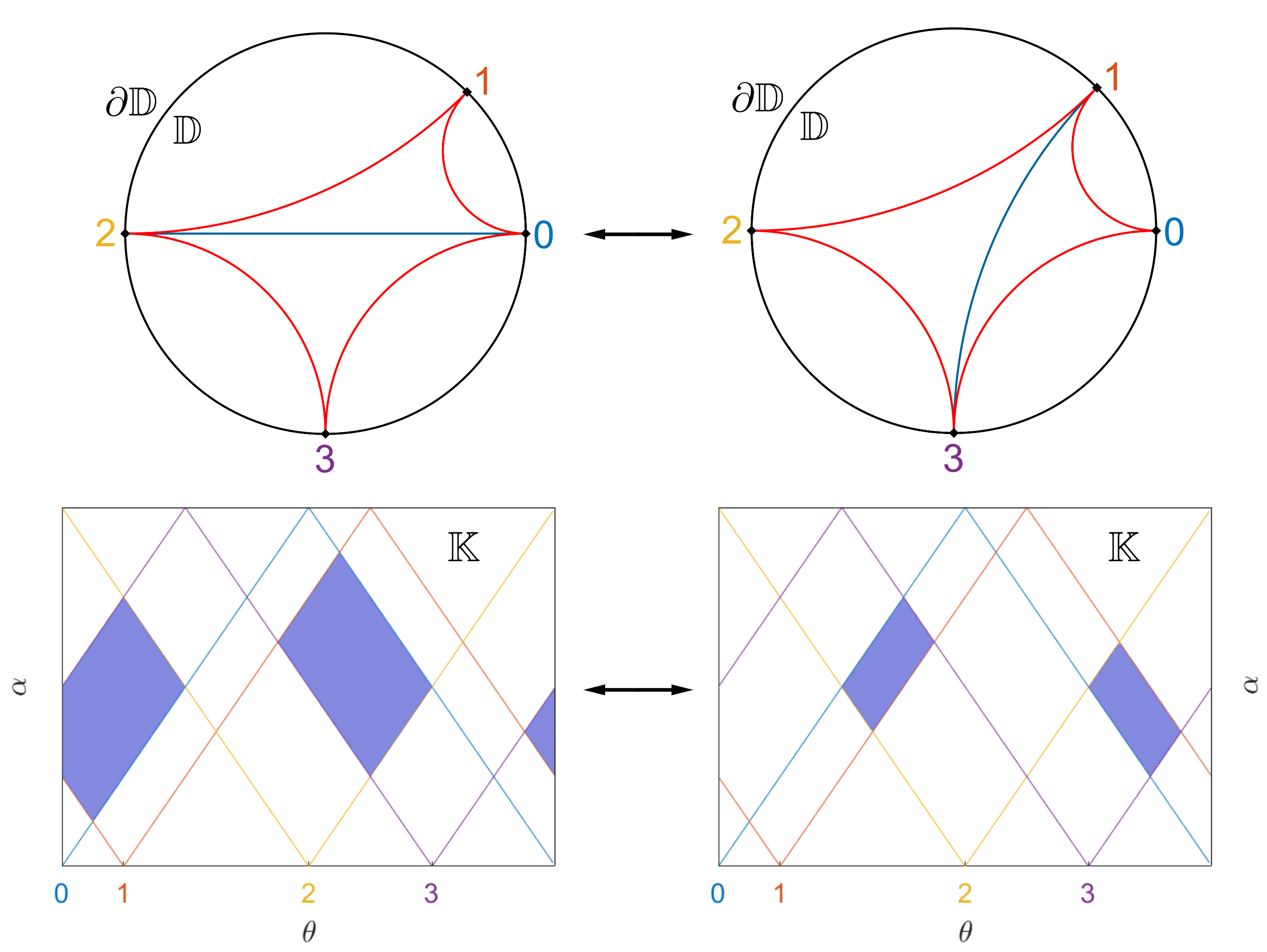}}
\caption{The triangulations of a geodesic quadrangle related by a flip in the bulk, and the associated causal patterns in kinematic space.}
\end{figure}

The quantity $u$ is associated to the diagonal connecting the points $0$ and $2$.
Clearly for the diagonal connecting the ones $1$ and $3$, we can define $t_{01,23}=t_{23,01}=1/u$. 
In this notation we have
\begin{subequations}
\begin{align}
I(12,30\vert 23)&=\frac{\mathfrak{c}}{3}\log(1+t_{12,30})=\frac{\mathfrak{c}}{3}\log(1+u),\\
I(01,23\vert 12)&=\frac{\mathfrak{c}}{3}\log(1+t_{01,23})=
\frac{\mathfrak{c}}{3}\log\left(1+\frac{1}{u}\right)
\label{quadentrop}
\end{align}
\end{subequations}
where entries like $12$ indicate the corresponding boundary region, i.e. the one between the points $1$ and $2$.
Notice that for a cyclic (counter clock wise) orientation of $\{0,1,2,3\}$ the $t$s are positive. According to Figure 2. their logarithms describe shears.

Now for a geodesic $N$-gon with a choice for its $n=N-3$ diagonals (giving rise to a triangulation), 
we define for $j\in\mathbb Z$ and $k=1,2,\dots ,n$
a special set of variables
\begin{equation}
\frac{1}{X_{j,k}}=1+t_{ab,cd},\qquad j+k\equiv 0\quad {\rm mod}2
\label{cimkezes}
\end{equation}
where
\begin{equation*}
b\equiv\frac{j-k}{2},\quad c\equiv\frac{j+k}{2},\quad a\equiv b-1,\quad d\equiv c+1\quad{\rm mod}(n+3)
\end{equation*}

\begin{figure}
{\includegraphics[height=3.5truecm]{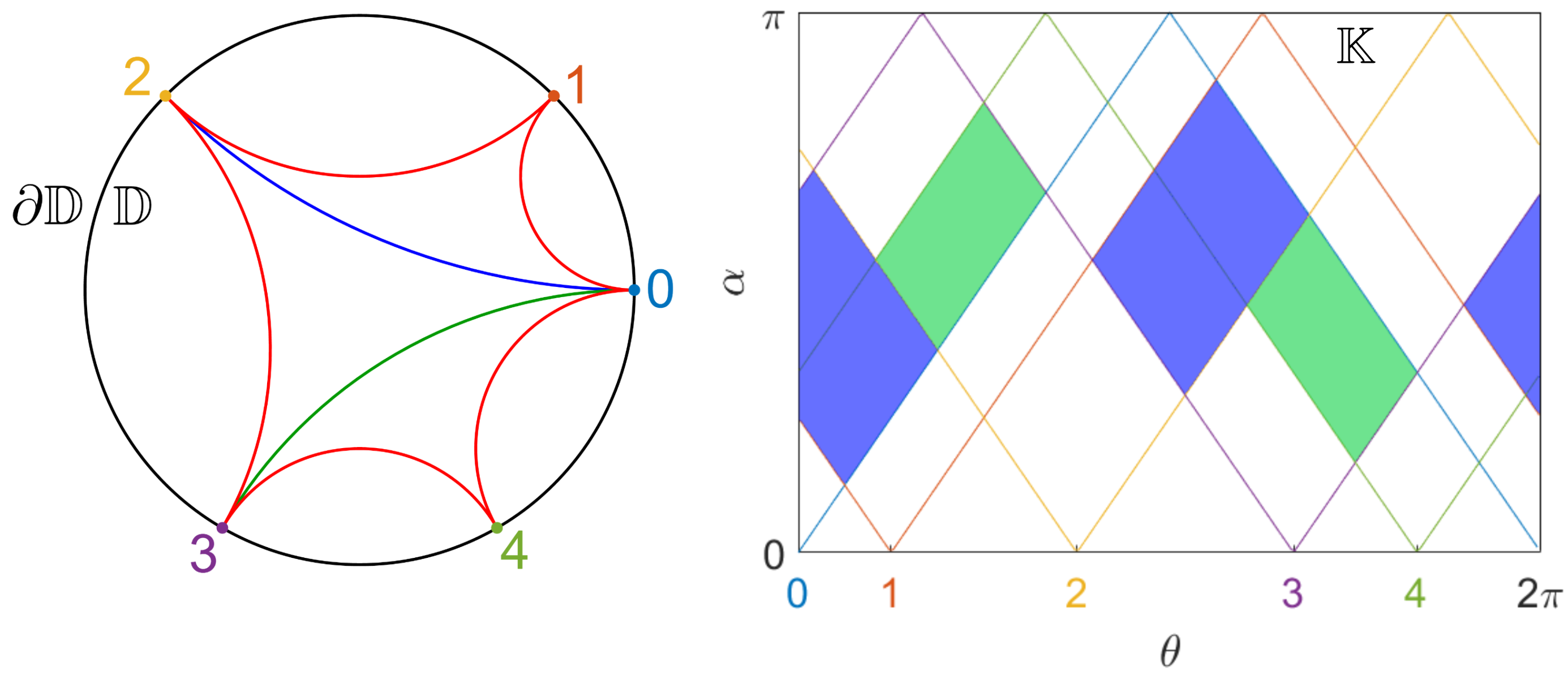}}
\caption{A triangulation of a geodesic pentagon, and the associated causal pattern in kinematic space.
The causal pattern is the one answering the boundary binary bracketing of regions of the form $(((A_0A_1)A_2)A_3)$.
See Section VI. for notation.}
\end{figure}

This special set of variables defines a special set of conditional mutual informations
\begin{equation*}
I_{j,k}\equiv I\left(ab,cd\vert bc\right),\qquad j\in\mathbb Z,\quad k=1,\dots,n
\end{equation*}
where Eqs. (26a-b) show that
\begin{equation}
X_{j,k}=e^{-\frac{3}{\mathfrak{c}}
I_{j,k}},\qquad j+k\equiv 0\quad{\rm mod}2.
\label{lenyeg}
\end{equation}

For example for a geodesic pentagon $N=5$, $n=2$ with a cyclic labelling $\{0,1,2,3,4\}$ we choose the triangulation by considering the diagonals $03$ and $02$. They are featuring the geodesic quadrangles $2340$ and $1230$.
We associate to them the positive quantities $u_1=t_{40,23}=t_{23,40}$ and $u_2=t_{30,12}=t_{12,30}$.
Then with this choice of triangulation and values for the pair $(u_1,u_2)$ 
we have
\begin{subequations}
\begin{align}
\frac{1}{X_{1,1}}&\equiv 1+\frac{u_1u_2}{1+u_1}=1+t_{40,12}\\
\frac{1}{X_{2,2}}&\equiv 1+u_1=1+t_{40,23}\\
\frac{1}{X_{3,1}}&\equiv 1+\frac{1}{u_2}=1+t_{01,23}\\
\frac{1}{X_{4,2}}&\equiv 1+\frac{1}{u_1(1+u_2)}=1+t_{01,34}\\
\frac{1}{X_{5,1}}&\equiv 1+\frac{u_2}{1+u_1+u_1u_2}=1+t_{12,34}
\end{align}
\end{subequations}
Notice that we can extend this set by adjoining to them new plaquette variables satisfying the formula
\begin{equation}
X_{j+5,3-k}=X_{j,k},\qquad j=0,1,2,3,4.
\label{adj}
\end{equation}
Now according to Eq.(\ref{integral}), the logarithms of these quantities are proportional to the areas of the regions in $\mathbb K$ that are bounded by point curves labelled by the numbers $a,b,c,d\in\{0,1,\dots N-1\}$ showing up in $t_{ab,cd}$.
More precisely, according to Eq.(\ref{Crofton}) $\frac{c}{3}$ times these logarithms give the areas of such regions with respect to the Crofton form. These areas in kinematic space encode conditional mutual informations of the corresponding regions in $\partial\mathbb D$.
Notice that due to the periodicity of kinematic space, by also using the adjoint set of variables of Eq.(\ref{adj}) no more $X_{j,k}$ plaquett variables are needed. Indeed, the central belt of $\mathbb K$ is already covered by them. This is reflected in the periodicity property $X_{j+10,k}=X_{j,k}$.

\begin{figure}
{\includegraphics[height=5.5cm]{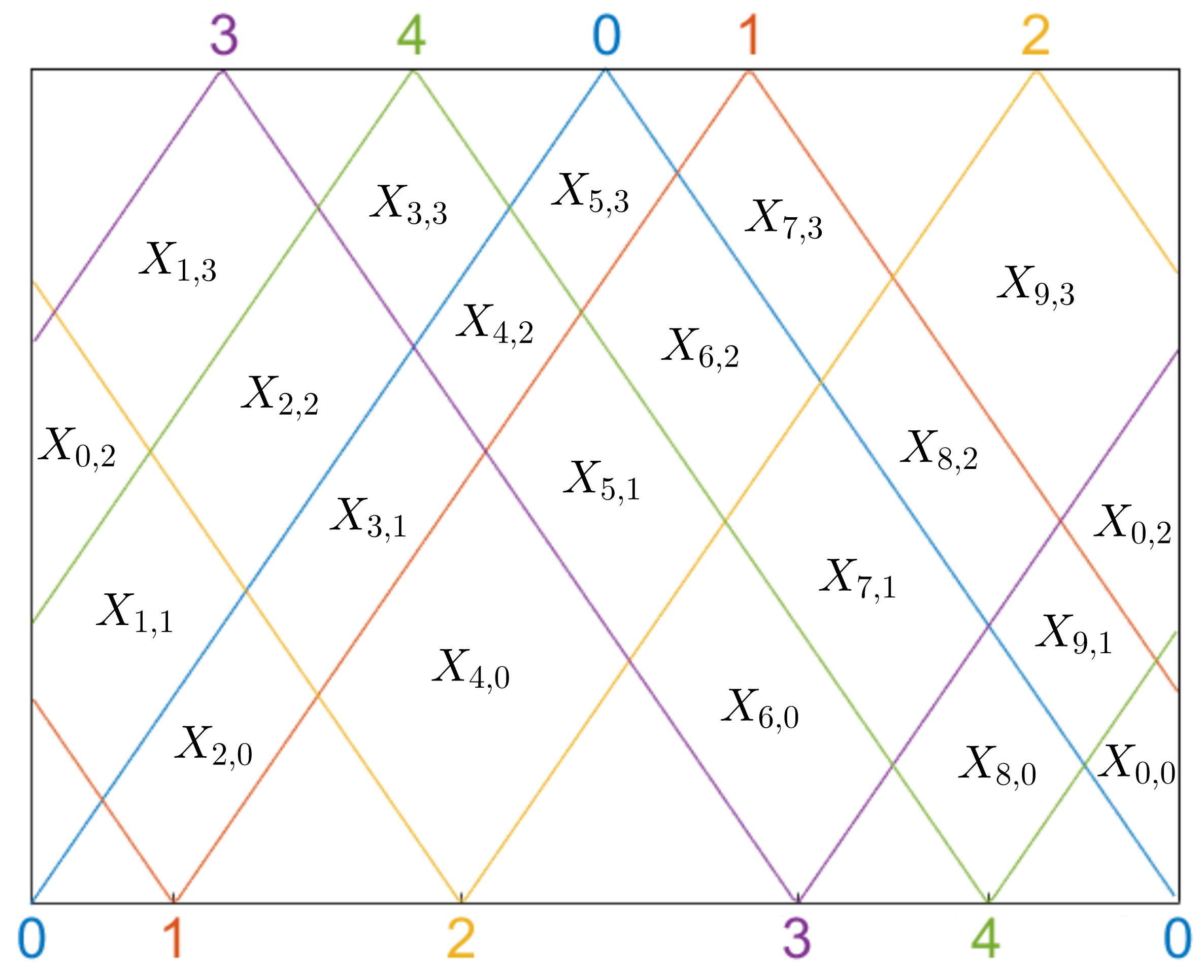}}
\caption{Cluster labelling of the kinematic space for a geodesic pentagon of Figure 6.}
\end{figure}

As an example one can consider the pair of variables $(X_{5,1},X_{6,2})=(X_{5,1},X_{1,1})$. According to Figures 6. and 7. taken together the corresponding two regions cover the region between the point curves of the boundary points $1,2$ and $3,0$. Taking the logarithm of
$\left(X_{5,1}X_{6,2}\right)^{-1}$ which is
\begin{equation}
\left(1+\frac{u_2}{1+u_1+u_1u_2}\right)\left(1+\frac{u_1u_2}{1+u_1}\right)=1+u_2
\end{equation}
and recalling that $u_2=t_{12,30}$ illustrates the additivity of conditional mutual information\cite{Czech1}, i.e.
$I(12,34\vert 23)+I(12,40\vert 24)=I(12,30\vert 23)$.
This reproduces a well-known algebraic property of these measures of entanglement.

 However, one can also reveal a more intricate algebraic structure of these measures.
In order to elaborate on this structure
it is worth introducing new variables $X_{j,0}$ and $X_{j,n+1}$. They will correspond to regions extending to the conformal boundary of $\mathbb K$. 
Indeed, geodesics with zero opening angle correspond to points on $\partial\mathbb D$, and such geodesics are located at the conformal boundary of kinematic space\cite{Zuk}. Now unlike the $X_{j,k}$s with $k=1,2,\dots n$ encoding conditional mutual informations like $I(A,C\vert B)$, the new variables are encoding {\it mutual informations} of the form $I(A,B)=S(A)-S(AB)+S(B)$.
However, since these regions are extending to the conformal boundary of $\mathbb K$, they have diverging areas with respect to the Crofton form.
Notice that this is as it should be since, unlike the $I(A,C\vert B)$s, the $I(A,B)$s are cutoff dependent (divergent) quantities of $\partial\mathbb D$.
The resulting divergence can explicitly be seen for example by taking the limit $\varphi_3\to\varphi_2$ in Eq.(\ref{teshear}) yielding a diverging $t_{12,30}$. 
Now due to (\ref{cimkezes}) such limiting cases yield vanishing values for these new variables: $X_{j,0}=X_{j,n+1}=0$. In the following we implement these constraints as boundary conditions for the variables $X_{j,k}$.

Now the important observation we would like to make is that for the (\ref{lenyeg}) set of conditional mutual informations
under the boundary conditions
\begin{equation}
X_{j,0}=X_{j,n+1}=0
\label{bcond}
\end{equation}
the following recursion relation holds
\begin{equation}
\left(1-\frac{1}{X_{j+1,k}}\right)\left(1-\frac{1}{X_{j-1,k}}\right)=\left(1-X_{j,k+1}\right)\left(1-X_{j,k-1}\right).
\label{iksz}
\end{equation}

It is well-known that one can explicitly solve this recursion relation \cite{FrenkelSzenes}.
The solution rests on choosing a special set of $X_{j,k}$s which we call a {\it seed set}.
In the notation of Ref.\cite{FrenkelSzenes} let us define $a_k\equiv X_{k,k}$ where $k=1,\dots,n$ with the boundary condition of Eq.(\ref{bcond}) for $j=0$ and $j=n+1$ translated into $a_0=a_{n+1}=0$. 
Then
\begin{equation}
X_{j+2,j}=\frac{1-a_1a_2\cdots a_j}{1-a_1a_2\cdots a_{j+1}}
\label{fs1}
\end{equation}

\begin{equation}
X_{n-j-1,n-j+1}=\frac{1-a_na_{n-1}\cdots a_{n+1-j}}{1-a_na_{n-1}\cdots a_{n-j}}.
\label{fs2}
\end{equation}
Using this solution one can {\it prove} that the solution of the (\ref{iksz}) system is {\it periodic} in $j\in\mathbb Z$, i.e. one can prove that
\begin{equation}
X_{j+n+3,n+1-k}=X_{j,k}
\label{mobius}
\end{equation}
(see also Eq.(\ref{adj}) for the $n=2$ case) which yields $X_{j+2N,k}=X_{j,k}$ with $N=n+3$.

Notice that for geodesic $N$-gons the number of nonzero seed variables equals $n$ which is the number of diagonals of a triangulations.
Hence choosing a particular triangulation of a geodesic $N$-gon one can derive a set of nonzero seed values $a_k, k=1,2,\dots n$.
Using then the explicit solution of (\ref{iksz}) one can {\it generate} the conditional mutual information labels for all of the causal diamonds covering the central belt of $\mathbb K$. In this way for a particular triangulation one obtains a {\it labelled tiling} of kinematic space by patterns of entanglement of the CFT vacuum.
For example for the $N=5$, $n=2$ case after choosing the triangulation shown in Figure 6. with the associated shear coordinates $(u_1,u_2)$ the seed variables are $X_{1,1}$ and $X_{2,2}$ of 
Eq. (29a-b). These are the plaquette variables for the leftmost side of the central belt shown in Figure 7.
Then using Eqs.(\ref{fs1})-(\ref{fs2}) and the periodicity property one can generate all the variables for the central belt.

As another example one can take the $N=6$, $n=3$ case of the {\it regular} hexagon with the choosen triangulation given by the one of Figure 8. The shear coordinates associated to the corresponding diagonals are
\begin{equation}
(u_1,u_2,u_3)=(t_{02,34},t_{12,40},t_{24,50})
\end{equation}
These coordinates give rise to the seed variables 
\begin{equation*}
a_1= X_{1,1}=\frac{1+u_2}{1+u_2+u_2u_3}
\end{equation*}
\begin{equation*}
a_2=X_{2,2}=\frac{1+u_1+u_1u_2}{(1+u_1)(1+u_2)}
\end{equation*}
\begin{equation*}
a_3=X_{3,3}=\frac{1+u_1}{1+u_1+u_1u_2}
\end{equation*}
%Figure 8. helyett Figure 9.
In this case the seed corresponds to the three causal wedges in between the two blue lines in the leftmost part of the central belt of Figure 9. From this seed by a straightforward exercise one can generate all the conditional mutual information labels for the remaining causal diamonds.
\begin{figure}
{\includegraphics[width=\columnwidth]{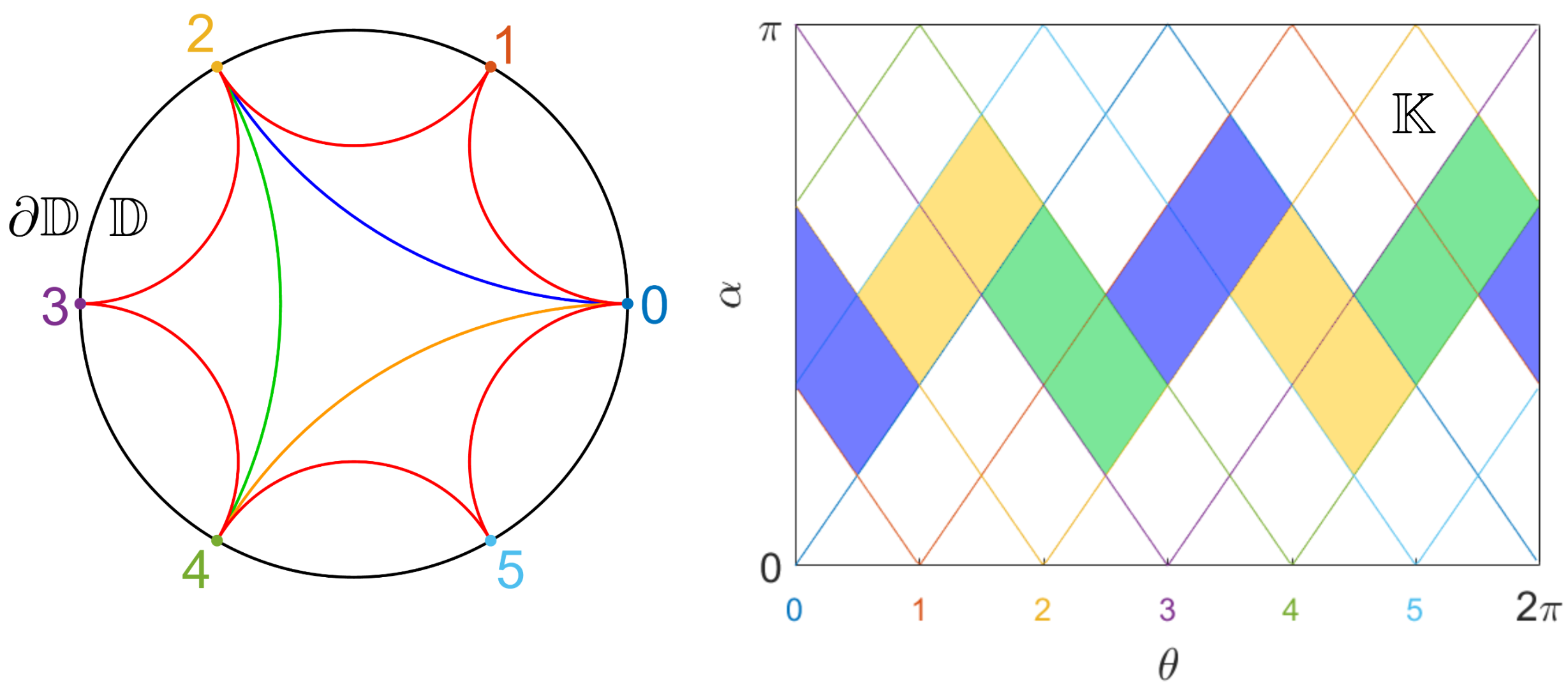}}
\caption{A triangulation of a regular geodesic hexagon, and the associated causal pattern in kinematic space.
The causal pattern is the one answering the boundary binary bracketing of regions of the form $(((A_0A_1)(A_2A_3))A_4)$. For notation see section VI.}
\end{figure}

%Figure 8 helyett Figure 9
It is important to realize that for {\it regular} $N$-gons (see Figure 9. for the $N=6$ case) the conditional mutual informations $I_{j,k}$ only depend on $k=1,2,\dots,n$. Indeed since the (\ref{kinmetric}) metric on kinematic space depends merely on $\alpha$ the areas of the causal diamonds in the regular $N$-gon case are depending merely on the discretized value of $\alpha$ which is the label $k$.
One can also see this from the explicit form of the (\ref{teshear}) shear coordinate which (in an obvious notation) gives
\begin{equation}
t_{j,k}=\frac{\sin^2\kappa}{\sin((k+2)\kappa)\sin(k\kappa)},\qquad \kappa=\frac{\pi}{N}
\end{equation}
a quantity independent of $j$ which is the discretization of the coordinate $\theta$.
This result gives the explicit form of the $j$ independent conditional mutual informations for regular $N$-gons
\begin{equation}
I_{j,k}=\frac{\mathfrak{c}}{3}\log\frac{\sin^2((k+1)\kappa)}{\sin^2((k+1)\kappa)- \sin^2\kappa},\quad k=1,\dots,n.
\end{equation}
Employing now the basic relation of Eq.(\ref{lenyeg}) one obtains for the $j$ independent plaquette variables
\begin{equation}
X^{(0)}_{j,k}\equiv 1-\frac{\sin^2\kappa}{\sin^2((k+1)\kappa)}
\label{jeindep}
\end{equation}
familiar from Ref.\cite{FrenkelSzenes}

Let us finally connect our patterns of entanglement to some structures well-known in the literature.
First of all notice that the $j$ independent expression of Eq.(\ref{jeindep}) solves our basic equation of 
Eq.(\ref{iksz}). The explicit form of the $j$ independent version of this equation can be written in the form
\begin{equation}
X_{j,k}^2=\prod_{l=1}^n\left(1-X_{j,k}\right)^{A_{kl}}
\end{equation}
where $A_{kl}$ is the Cartan matrix of the Dynkin diagram of type $A_n$.

Generally let us define
\begin{equation}
Y_{j,k}=\frac{X_{j,k}}{1-X_{j,k}}=\frac{1}{e^{\frac{3}{\mathfrak{c}}I_{j,k}}-1}
\label{ipszilon}
\end{equation}
where $j\in{\mathbb Z}$ and $k=1,2,\dots,n$. Then the new form of the (\ref{iksz}) recursion relation is
\begin{equation}
Y_{j-1,k}Y_{j+1,k}=(1+Y_{j,k-1})(1+Y_{j,k+1})
\label{zamoan}
\end{equation}
with boundary conditions $Y_{j,0}=Y_{j,n+1}=0$.
This recursion is defining a Zamolodchikov Y-system of type $(A_n,A_1)$ \cite{Zamo}. It is known that any solution of the Thermodynamic Bethe Ansatz equations (TBA) for a certain type of theories labelled by pairs $(G,H)$ of Dynkin diagrams of ADE type satisfies a set of equations which boils down to a recursion generalizing the (\ref{zamoan}) form \cite{Ravanini}. 

The result\cite{FrenkelSzenes,Gliozzi} 
\begin{equation}
Y_{i+n+3,n+1-j}=Y_{i,j}
\label{yperiod}
\end{equation}
is called Zamolodchikov Periodicity. In our treatise it shows up naturally in the scanning process of 2D de Sitter spacetime (now playing the role as the kinematic space $\mathbb K$) by entanglement patterns.
In closing this section let us summarize the basic relations between the plaquaette variables $X_{j,k}$ and $Y_{j,k}$
and the shear coordinate $t_{j,k}$
\begin{equation}
\frac{1}{X_{j,k}}=1+\frac{1}{Y_{j,k}}=1+t_{j,k}.
\label{mnemonic}
\end{equation}

\section{Duality}

We have seen that our encoding of entanglement patterns into kinematic space was based on a Zamolodchikov $Y$-system
of type $(A_n,A_1)$. The structure of $Y$ systems of type $(G,H)$ is determined by the adjacency structure of the corresponding Dynkin diagrams. Now the set of vertices ${\mathcal I}$ of any Dynkin diagram is a disjoint union of two sets ${\mathcal I}_+$ and ${\mathcal I}_-$  such that there is no edge between any two vertices of ${\mathcal I}_+$ nor between any two vertices of ${\mathcal I}_-$. 
For the Dynkin diagram of type $A_n$ this means that for ${\mathcal I}\simeq\{1,2,\dots,n\}$ we have two types of vertices labelled by even or odd numbers.
Let us then define for $k=1,2,\dots,n$
\begin{equation*}
\epsilon(k)=\begin{cases} +1,&\text{if $k$ is even}\\
-1,&\text{if $k$ is odd}\end{cases} 
\end{equation*}
and let
\begin{equation}
b_{kl}=2\delta_{kl}-A_{kl}
\end{equation}
the adjacency matrix of the Dynkin diagram of type $A_n$ where $A_{kl}$ is the corresponding Cartan matrix.

Notice now that for $j\in{\mathbb Z}$ and $k=1,2,\dots n$ the left hand side of Eq.(\ref{zamoan}) has a fixed parity $\epsilon(k)(-1)^j$.
Therefore our $Y$ system decomposes into two independent systems, an even one and an odd one.
Until this point we have been considering merely the one of even type corresponding to our choice of $j+k\equiv 0$ mod $2$ see Eqs.(\ref{lenyeg}) and (\ref{ipszilon}).
Now the question arises: What kind of patterns of entanglement are encoded into kinematic space by systems of the odd type?

Without the loss of generality one may assume that\cite{Williams}
\begin{equation}
Y_{j+1,k}=Y^{-1}_{j,k}, \text{whenever $\epsilon(k)=(-1)^j$}
\label{transition}
\end{equation}
and as a result of this Eq.(\ref{zamoan}) can be rewritten
\begin{equation*}
Y_{j+1,k}=\begin{cases}
Y_{j,k}\prod_{l=1}^{n}\left(1+Y_{j,l}\right)^{b_{kl}},&\text{if $\epsilon(k)=(-1)^{j+1}$}\\
Y^{-1}_{j,k},&\text{if $\epsilon(k)=(-1)^{j}$}\end{cases}
\end{equation*}

\begin{figure}
{\includegraphics[height=5.5cm]{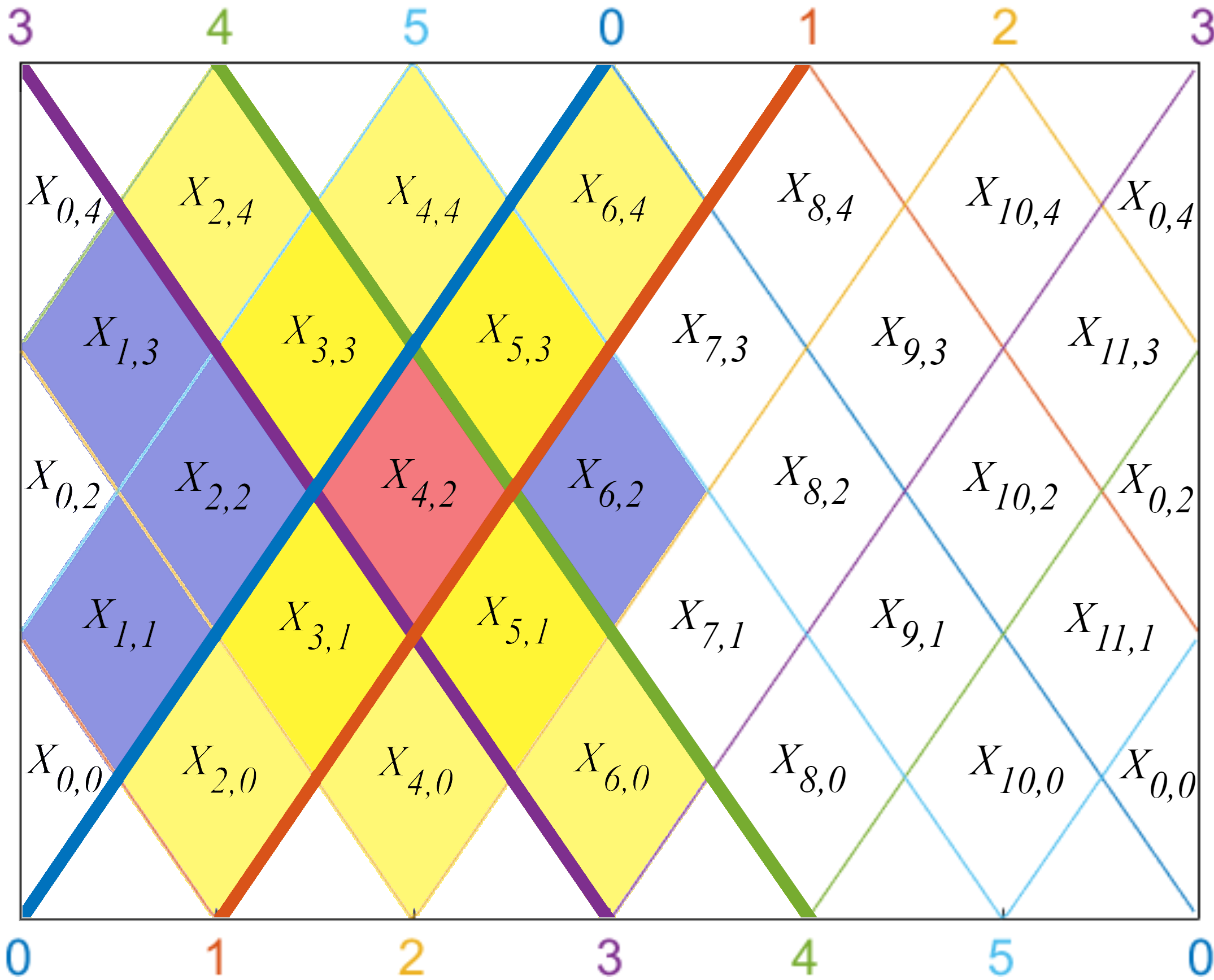}}
\caption{Cluster labelling of kinematic space for a regular geodesic hexagon. The light cone of the variable $X_{4,2}$ highlighted in red is shown. The causal diamonds inside and on its light cone are the yellow regions, and the diamonds outside are the blue ones. Due to Zamolodchikov periodicity the causal diamonds belonging to a fundamental domain are coloured with blue and dark yellow. Notice that merely the blue diamonds ($\overline{C_{4,2}}$) and the red one of the fundamental domain belonging to the central belt of $\mathbb K$ are featuring Eq.(\ref{belt}).}
\end{figure}

As an illustration of this result for geodesic pentagons ($N=5$,$n=2$) one can use this recursion with the initial values 
\begin{equation}
Y_{0,1}=v_1\equiv \frac{1}{u_1(1+u_2)},\qquad
Y_{0,2}=v_2\equiv \frac{1+u_1+u_1u_2}{u_2}
\label{kuksi}
\end{equation}
to obtain the set
of labels featured in Figure 10. One should start with the lower right part of the diagram having the green label of Eq.(\ref{kuksi}) and proceed in a counter clock-wise fashion.
In Figure 10. the green labels are of even and the red ones are of odd type. By virtue of Eq.(\ref{mnemonic}) the green labels then reproduce the $X_{j,k}$ variables of Eqs.(29) of even type.
(In arriving at this result according to Eq.(\ref{yperiod}) $Y_{0,2}=Y_{5,1}$.)
Notice also that using our recursion after completing a full counter clock-wise circle in Figure 10. the green and red labels are exchanged.
This is in accord with Zamolodchikov periodicity, since the initial labels are restored after {\it two such cycles}, i.e. one needs ten flips to get back.
Recall also that this periodicity is just the discretized version of the periodicity of kinematic space in the $\theta$ coordinate.

In order to clarify the meaning of the "even-odd" transition of Eq.(\ref{transition}), let us first record the obvious fact that it is related to flips of the corresponding quadrangles (see Figure 10.).
Next one should notice that for $\epsilon(k)=(-1)^j$ a transition of the type $Y\mapsto 1/Y$ corresponds to a one $X\mapsto 1-X$.
Now by plugging $j=n$ in Eq.(\ref{fs1}) one obtains $X_{n+3,n}=1-X_{n+2,n}=\prod_{k=1}^nX_{k,k}$. Hence the odd variable $X_{N,n}$ can be written as a product of even ones.
However, one can even prove more. For example for the $N=6, n=3$ case apart from the corresponding relation $X_{6,3}=X_{1,1}X_{2,2}X_{3,3}$ a set of relations of more general kind hold. Indeed one can check for example that
\begin{equation}
X_{5,2}=1-X_{4,2}=X_{1,1}X_{2,2}X_{1,3}X_{6,2}
\label{belt}
\end{equation}
and from Figure 9. one discovers that the labels showing up in the product are the ones lying in the complement of the light cone of $X_{4,2}$. More precisely since $X_{4,2}=X_{10,2}$ and the complements of the light cones are overlapping, if we avoid double counting the labels are {\it precisely} the ones showing up in the complement of the light cone of $X_{4,2}$ in the central belt of $\mathbb K$.
In fact our example is just a special case of Proposition 4. of Ref.\cite{FrenkelSzenes}
\begin{equation}
1-X_{j,k}=\prod_{(j^{\prime},k^{\prime})\in \overline{\mathcal C}_{j,k}}X_{j^{\prime},k^{\prime}}
\label{kupos}
\end{equation}
where $1\leq j\leq n+3$ and $1\leq k\leq n$ , $j+k\equiv 0$ mod$2$ and
\begin{equation*}
\overline{\mathcal C}_{j,k}=\{(j^{\prime},k^{\prime})\vert \vert k^{\prime}-k\vert <\vert j^{\prime}-j\vert\}.
\end{equation*}
Since $1-X_{j,k}=X_{j+1,k}$ and the {\it negatives} of the logarithms of the $X$s are related to conditional mutual informations one can then {\it define} the $I_{j+1,k}$s as quantities related to a {\it dual system} of $X$ variables also satisfying the (\ref{iksz}) recursion relation.

\begin{figure}
{\includegraphics[width=\columnwidth]{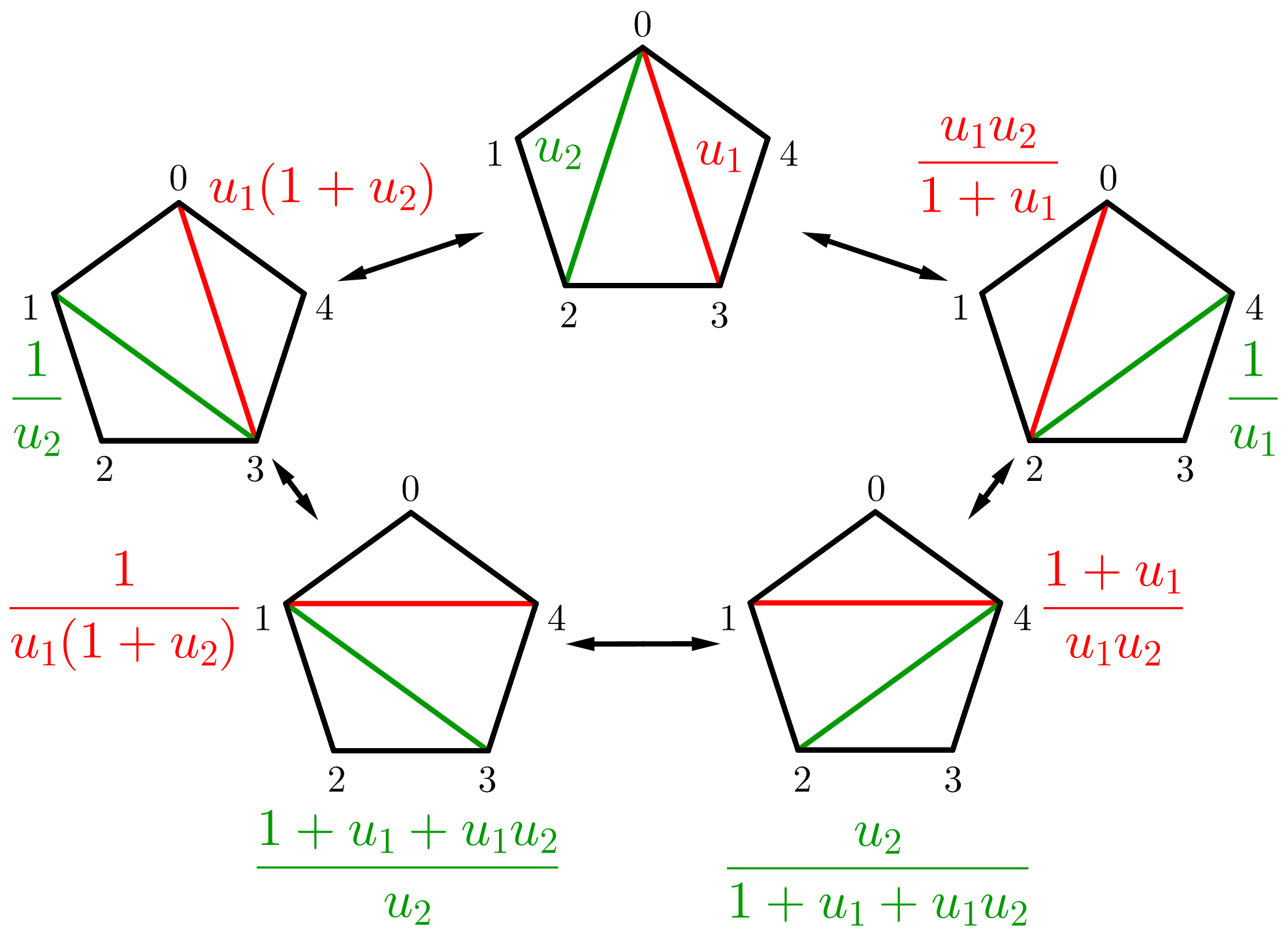}}
\caption{Triangulations of a geodesic pentagon, and the associahedron ${\mathcal K}^2$.
Here the labels are shear coordinates $t_{j,k}$ related to the variables $X_{j,k}$ and $Y_{j,k}$ by Eq.(\ref{mnemonic}). 
The green labels give rise to the explicit expressions of the even variables $X_{j,k}$ with $j+k\equiv 0$ mod $2$ familiar from Eq.(29). The red ones give rise to the corresponding  odd variables with $j+k\equiv 1$ mod $2$.}
\end{figure}

Using this result and Eqs.(\ref{lenyeg}), (\ref{mnemonic}), (\ref{otherdist}) and the Brown-Henneaux relation one obtains
\begin{equation}
\frac{\ell_{\rm AdS}}{G_{\rm N}}\cdot\log\sinh\frac{\ell_{j,k}}{2}=I_{j,k}-\sum_{(j^{\prime},k^{\prime})\in \overline{\mathcal C}_{j,k}}I_{j^{\prime},k^{\prime}}.
\label{huha}
\end{equation}
Here $\ell_{j,k}$ is the distance in the bulk between the the geodesics anchored to the boundary regions $[(j-k)/2,(j+k)/2]$ and  $[(j-k-2)/2,(j+k+2)/2]$. 
This equation shows how the difference between the conditional mutual informations of even ($I_{j,k}$) and odd type ($I_{j+1,k}$) of the boundary are related to the corresponding geodesic lengths $\ell_{j,k}$ of the bulk, or alternatively to the proper times $\Delta\tau_{j,k}$ elapsed between the two time-like separated vertices of the relevant causal diamonds in $\mathbb K$ (see the right hand side of Figure 4.).
Notice that the alternative expression valid for an arbitrary geodesic $N\geq 4$-gon
\begin{equation}
I_{j,k}-I_{j+1,k}=\frac{\mathfrak{c}}{3}\log t_{j,k}
\end{equation}
can be regarded as a generalization of Eq.(\ref{shearRT}) obtained for geodesic quadrangles.

These considerations show that for large $n$ and in terms of $X$ variables the $(A_n,A_1)$ type Zamolodchikov systems of even and odd kind have an interesting entanglement interpretation in kinematic space.
The $X_{j,k}$ of even systems are labelled by {\it points} $(j,k)$ in kinematic space and the odd ones with {\it complements of light cones} of such points. Correspondingly there are two possible ways of scanning kinematic space by patterns of entanglement provided by causal domains associated to the $X_{j,k}$s. The two different ways are related by duality between points and the complements of their light cones. 
In the following section we elaborate on this scanning process.

\section{Scanning Kinematic space}

We have seen that by discretizing $\partial\mathbb D$ to $N$ regions and considering the corresponding gauge invariant conditional mutual informations of the CFT vacuum, from a seed pattern we can recursively generate area labels for a corresponding discretization of $\mathbb K$. The starting point of this generating process was a choice of a geodesic $N$-gon taken together with a particular triangulation of the bulk $\mathbb D$. 
Now a natural question to be asked is: How this process of encoding entanglement patterns of the boundary into plaquette variables of $\mathbb K$ is depending on the choice of triangulations?

For a geodesic $N$-gon we label the $N$ points on the boundary $\partial\mathbb D$ by the numbers $\{0,1,\dots, N-1\}$.
These points partition the boundary into $N$ regions $A_0,A_1,\dots A_{N-1}$ where $A_j$ is a region between $j$ and $j+1$. Note that region $A_{N-1}$ between $N-1$ and $N\equiv 0$ mod $N$ is redundant since it is fixed by the list $A_0,A_1,\dots ,A_{N-2}$.
Given a geodesic $N$-gon a diagonal is a geodesic between two non-adjacent boundary points. A partial triangulation is a collection of mutually non-crossing diagonals. A full triangulation or simply a triangulation is a partial triangulation with maximal (i.e. $N-3$) number of diagonals.
It is known\cite{Lame} that for an $N\geq 3$-gon the number of (full) triangulations is given by the Catalan number $C_{N-2}=\frac{1}{N-1}\binom{2(N-2)}{N-2}$. Hence for $N=3,4,5,6,7,8\dots $ we get the numbers $C_{N-2}=1,2,5,14,42,132,\dots$.

A list ${\bf A}_0{\bf A}_1\dots {\bf A}_{N-1}$ will refer to the sequence of consecutive Ryu-Takayanagi geodesics of $\mathbb D$  anchored to the boundary regions $A_0A_1\dots A_{N-1}$ forming a geodesic $N$-gon.
In the following we adopt the convention of referring to a diagonal by introducing a bracketing to this list.
Hence for $N=4$ the symbol $({\bf A}_0{\bf A}_1){\bf A}_2{\bf A}_3$ is a path of consecutive geodesic segments forming a quadrangle taken together with a diagonal connecting the boundary points $0$ and $2$. 
Notice that since ${\bf A_3}$ is redundant, by an abuse of notation one can omit the last entry and use the binary bracketing $(({\bf A}_0{\bf A}_1){\bf A}_2)$ instead. Indeed, the outer extra bracket can be regarded as a reference to connecting the boundary points $0$ and $3$ which is just ${\bf A_3}$ elevated to the status of a "degenerate diagonal".
The net result is that either of the symbols: the shorter $({\bf A}_0{\bf A}_1){\bf A}_2$, or the longer one $(({\bf A}_0{\bf A}_1){\bf A}_2)$ can be regarded as a mnemonic for a geodesic quadrangle taken together with a diagonal connecting the boundary points $0$ and $2$.

In the boundary binary bracketings like $(A_0A_1)A_2$ or $((A_0A_1)A_2)$ has an obvious meaning related to the causal structure their corresponding regions enjoy.  
Indeed, the regions showing up in any meaningful binary bracketing are either time-like or light-like separated. We adopt the phrase that a set of regions is containing {\it compatible} pairs if and only if no two of such pairs are space-like separated.
In our example we have the regions $A_0,A_1,A_2$ and the combined region $A_0A_1$ no two of them are space-like separated. (An example for space-like separated regions is provided by the pair $(A_0A_1,A_1A_2)$.) The alternative notation $((A_0A_1)A_2)$ makes also reference to the extra region $A_3$. Of course its inclusion into the list of regions is not destroying the compatibility structure.

To any meaningful binary bracketing (containing compatible pairs of regions) of the boundary one can associate a triangulation of the bulk. The corresponding binary bracketing of the list of consecutive geodesics  ${\bf A}_0{\bf A}_1\dots {\bf A}_{N-1}$ will be regarded as a representative of the particular triangulation in $\mathbb D$.
Since geodesics are represented by points in kinematic space a binary bracketing for the list of symbols ${\mathcal A}_0{\mathcal A}_1\dots {\mathcal A}_{N-1}$ gives rise to a collection of points in $\mathbb K$ forming a {\it causal pattern}. For example it is easy to check that the bracketing $({\mathcal A}_0({\mathcal A}_1(\dots {\mathcal A}_{N-1})))$ gives rise to two triangular shapes contained in the upper and lower light cones of the point $N-1$. 
The points ${\mathcal A}_0{\mathcal A}_1\dots {\mathcal A}_{N-1}$ are light-like separated and they are lying on the right hand side of the corresponding light cones. 
An illustration of this for $N=6$ can be seen in the top pattern of Figure 13.
Further examples for causal patterns can be seen in Figures 6. and 8. In the first case the causal pattern is the one answering the boundary binary bracketing of regions of the form $(((A_0A_1)A_2)A_3)$, and in the second case of the form $(((A_0A_1)(A_2A_3))A_4)$.

For $N=4$ the two possible bulk triangulations are shown in Figure 5. In the boundary using the binary bracketing one can refer to them as: $((A_0A_1)A_2)$ and $(A_0(A_1A_2))$. We also know that to the two possible diagonals 
one can associate shear coordinates $u=t_{12,30}$ and $1/u$ which in turn encode the conditional mutual informations of Eqs.(26a-b). A change from one of the diagonals of a particular geodesic quadrangle to the other one is a {\it flip}. Since diagonals are taken together with shear labels we see that for the $N=4$ case  a flip is also associated with the change $u\mapsto 1/u$. In terms of the conditional mutual informations related by flips the shear is expressed by Eq.(\ref{shearRT}). 
In the following we will call the flips of causal patterns answering the flip in the corresponding triangulation a {\it mutation of a causal pattern}. An example for a mutation of a causal pattern can be seen in the lower part of Figure 5.

Now for $N\geq 3$ one can define a convex polytope of dimension $N-3$ called the {\it associahedron} ${\mathcal K}^{N-3}$, such that for every $d=0,1,\dots N-3$ there is a one-to-one correspondence between its $d$-dimensional boundaries and the $d$-diagonal partial triangulations of the geodesic $N$-gon\cite{Stasheff,Nima1}. Here by boundary we mean a boundary of the polytope of any codimension.
Moreover, a codimension $d$ boundary $B_1$ and a codimension $d+l$ boundary $B_2$ are adjacent if and only if the partial triangulation of $B_2$ can be obtained by addition of $l$ diagonals to the partial triangulation of $B_1$.
The vertices of the associahedron correspond to the (full) triangulations. The labeling of the vertices of ${\mathcal K}^2$ and ${\mathcal K}^3$ by such triangulations can be seen in Figures 10. and 11. Notice that adjacent triangulations are related by a flip.   
The associahedron boundaries related to partial triangulations are illustrated for ${\mathcal K}^3$ in Figure 12.

\begin{figure}
{\includegraphics[height=6cm]{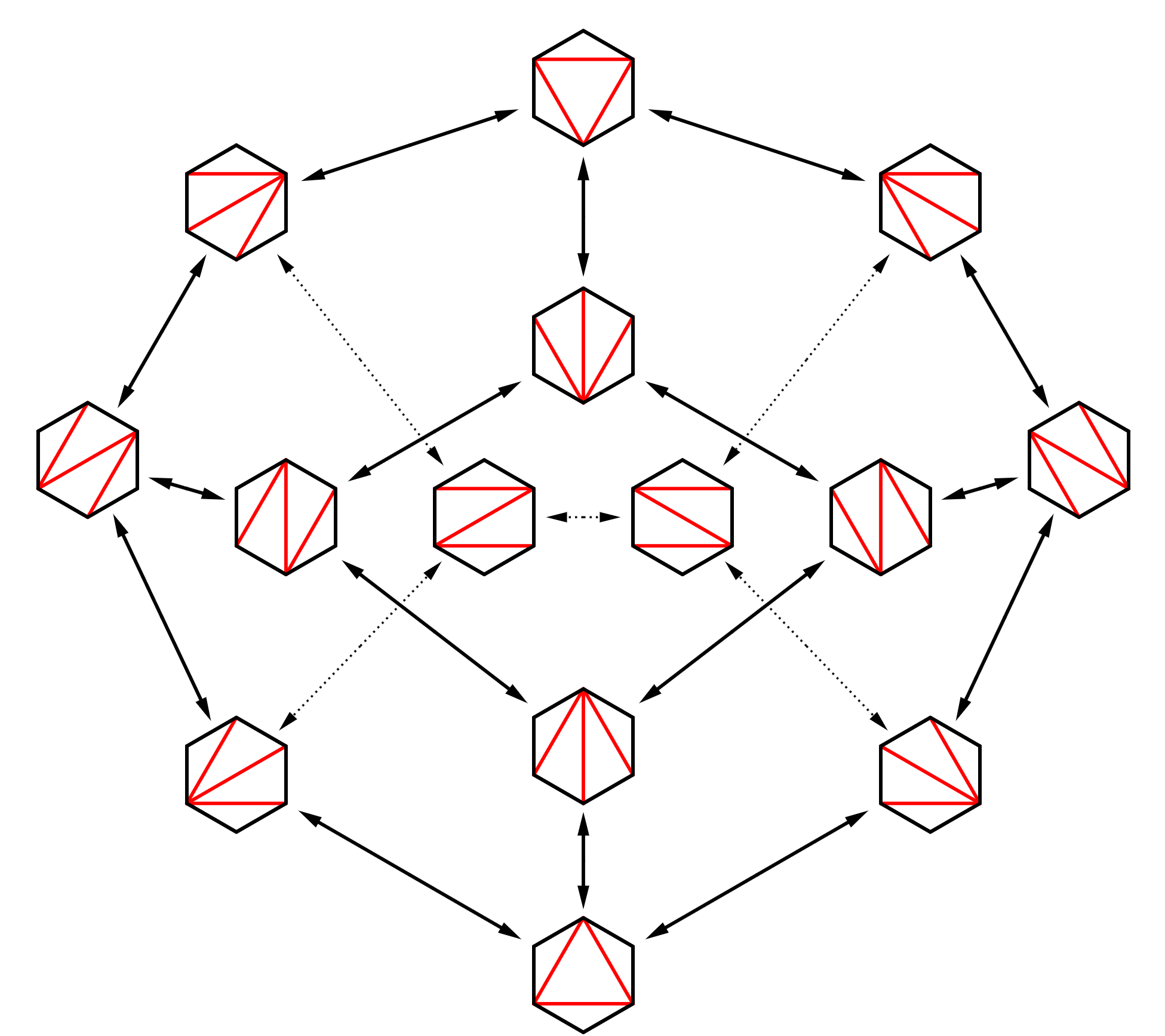}}
\caption{Triangulations of a geodesic hexagon, and the associahedron ${\mathcal K}^3$. The vertices of the associahedron are labelled by triangulations of $\mathbb D$. Adjacent triangulations are related by a flip of one of the diagonals belonging to a geodesic quadrangle.}
\end{figure}

For $N=5$ the list of all possible binary bracketings of boundary regions are as follows
\begin{equation*}
((A_0(A_1A_2))A_3),\quad (((A_0A_1)A_2)A_3)\quad ((A_0A_1)(A_2A_3))
\end{equation*}
\begin{equation*}
(A_0((A_1A_2)A_3)),\quad (A_0(A_1(A_2A_3))).
\end{equation*}
The reader can check that the pentagonal-like arrangement of this list of binary bracketings encoding compatible boundary regions corresponds to the pentagonal arrangement of bulk triangulations of Figure 10.
Moreover, in kinematic space these triangulations give rise to five causal patterns. These patterns are arising from the middle top one triangulation of Figure 10. after representing it in $\mathbb K$, and then proceeding in a clock-wise fashion. Explicitly: it is easy to show that these patterns are arising from the one of Figure 6. by a cyclic shift to the left.
Then this sequence of mutations of causal patterns is scanning the central belt of kinematic space.

Let us now observe that the distribution of closing brackets uniquely determines the binary bracketing\cite{TamariHuang,Birkhoff,Geyer}, hence the causal pattern in $\mathbb K$.
Now if in the corresponding opening brackets we choose a constant distribution just one before each symbol $A_{j}, j=0,1,\dots N-2$ what we get is a {\it right bracketing}. The right bracketings in the $N=5$ case are as follows
\begin{equation*}
A_0(A_1(A_2))(A_3),\quad A_0(A_1)(A_2)(A_3)\quad A_0(A_1)(A_2(A_3))
\end{equation*}
\begin{equation*}
A_0(A_1(A_2)(A_3)),\quad A_0(A_1(A_2(A_3))).
\end{equation*}
Each right bracketing can be encoded into an $(N-2)$-vector $v\equiv (v_1,v_2,\dots,v_{N-2})$ as follows\cite{Birkhoff,Geyer}. For $\mu=1,2,\dots N-2$ we have $v_{\mu}=\nu$ if and only if the opening bracket before $A_{\mu}$ closes after $A_{\nu}$.
For example for the $N=5$ case we get the following set of $3$-vectors
\begin{equation*}
(2,2,3),\qquad (1,2,3),\qquad (1,3,3) 
\end{equation*}
\begin{equation*}
(3,2,3),\qquad (3,3,3).
\end{equation*}
Now one can prove\cite{TamariHuang,Birkhoff} that the components of our $(N-2)$-vectors, comprising a space ${\mathcal V}_{N-2}$, are positive integers $\leq N-2$ and they satisfy the following two constraints: $v_{\mu}\geq\mu$  and $v_{\mu}\geq \nu > \mu$ implies $v_{\mu}\geq v_{\nu}$ for $\mu=1,2,\dots N-2$. The latter condition captures the fact that two different pairs of brackets do not overlap, in other words no space-like separation of boundary regions is allowed. One can then prove that the number of such $(N-2)$-vectors is $C_{N-2}$, i.e. it equals the number of (full) triangulations of our geodesic $N$-gon. In this way we managed to characterize the space of causal patterns by a set of $(N-2)$-vectors of cardinality $C_{N-2}$.

\begin{figure}
{\includegraphics[height=6cm]{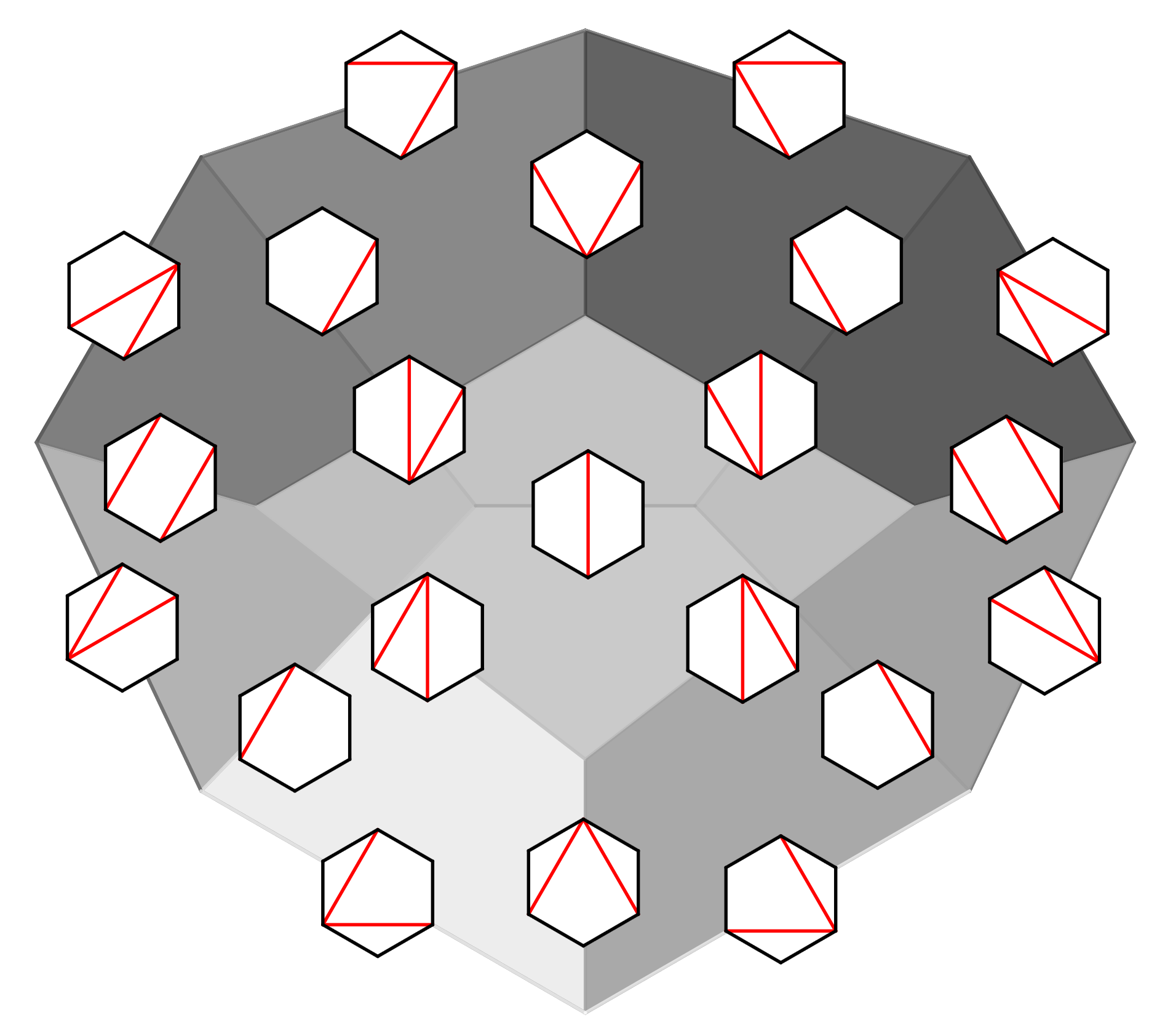}}
\caption{The associahedron ${\mathcal K}^3$ with its edges and faces labelled by partial triangulations of the geodesic hexagon. }
\end{figure}

Now the important point we would like to emphasize is that the space of causal patterns $T_{N-2}$ in $\mathbb K$ is also equipped with a partial order. Moreover, the corresponding partially ordered set $(T_{N-2},\leq)$ of causal patterns is isomorphic to a lattice: the Tamari lattice\cite{Tamari,TamariHuang}.
Indeed, let us identify the space of all binary bracketings with $T_{N-2}$. Then our $N-1$ boundary regions $A_j$ (recall that we have not included into the list the region $A_{N-1}$)) can be ordered according to the {\it semi-associativity rule}\cite{Tamari,TamariHuang,Geyer}: $(AB)C\mapsto A(BC)$.
Notice that as far as entanglement is concerned a particular bracketing of $N$ boundary regions can be interpreted as a choice of {\it context} in which our boundary subdivision should be considered. In this picture applying the semi-associative rule in $\partial\mathbb D$ is just the elementary change in the space of contexts. Notice that there is an ambiguity here in whether we choose the left or right ordering of brackets in the semi-associative rule.
Our choice is the one compatible with the rule represented in $\mathbb D$ by the usual flip in the corresponding geodesic quadrangle. Notice also that our flip in turn is represented by a causal flip, i.e. a mutation of a starting pattern to another one in $\mathbb K$. Now for two causal patterns $\mathcal{T}_1,\mathcal{T}_2\in T_{N-2}$ we define $\mathcal{T}_1\leq\mathcal{T}_2$ if and only if $\mathcal{T}_1$ can be transformed into $\mathcal{T}_2$ by repeated application of our semi-associativity rule.  
One can then identify $T_{N-2}$ with the set of $(N-2)$-vectors satisfying our aforementioned two constraints\cite{Birkhoff}, and the partial order in this set of vectors is just componentwise comparison. Hence for example for the $N=5$ case one has $(3,3,3)\geq (3,2,3)\geq (2,2,3)\geq (1,2,3)$ and $(3,3,3)\geq (1,3,3)\geq(1,2,3)$, however the vectors within the pairs  $((1,3,3),(3,2,3))$  and $((1,3,3),(2,2,3))$ are not related. 
These considerations establish the Tamari lattice structure for $T_3$, i.e. the space of causal patterns for $N=5$.

The Tamari lattice $T_4$ of causal patterns in $\mathbb K$ for $N=6$ can be seen in Figure 13.
Generally there is a maximal and minimal element of the lattice $T_{N-2}$. In the $(N-2)$-vector notation they correspond to the vectors
$(N-2,N-2,\dots,N-2,N-2)$ and $(1,2,\dots,N-1,N-2)$. In terms of Ryu-Takayanagi geodesics the corresponding diagonals are giving rise to "fans" emanating from the points $N-1$ and $0$ respectively. 
Clearly the corresponding causal patterns are just pairs of past/future light cones emanating from the points ${\mathcal A}_{N-1}\in\mathbb K$ or
${\mathcal A}_{0}\in\mathbb K$
lying inside the past/future light cones of the points (degenerate geodesics) $N-1$ and $0$ lying in past /future infinity of $dS_2$. 
For an illustration of this for $N=6$ see Figure 13.
Notice that each of the causal patterns is uniquely determined by $N-3$ points in $\mathbb K$ corresponding to the diagonals in $\mathbb D$. The reader can verify this statement by looking at the causal patterns and identifying their corresponding points in  Figures 6. and 8.  

In the following we adopt the convention of picturing these points inside the past light cone of the point $N-1$.
More precisely what we need is a part of this past light cone  which forms an isosceles right angular triangular region constructed as follows. Define the central belt of the grid, depicted for $N=5,6$ in Figures 7. and 9, as the region obtained by removing the plaquettes labelled by the variables $X_{j,0}$ and $X_{j,n+1}$ from the lattice. Then the triangular region is that part of the lattice with points lying inside and on the past light cone of ${\mathcal A}_{N-1}\in\mathbb K$ which has its base formed by the boundary of the central belt.

\begin{figure}
{\includegraphics[width=\columnwidth]{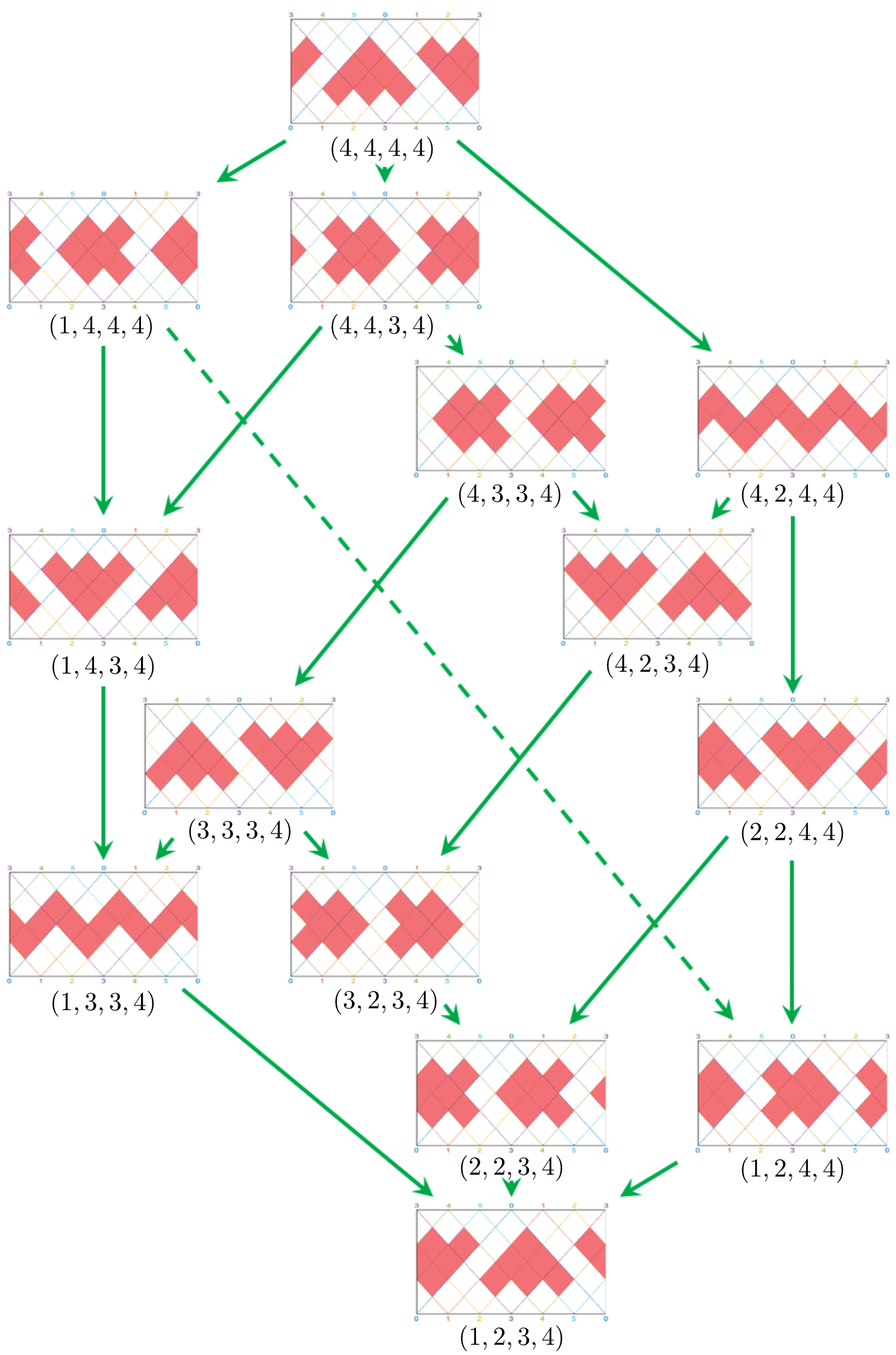}}
\caption{The Tamari lattice of causal patterns $(T_4,\leq)$ in kinematic space. Each causal pattern is labelled by a vector as an element of ${\mathcal V}_4$ defined in the text. In terms of these vectors the partial order on $T_4$
is represented by componentwise comparison. Notice that the causal patterns are showing up in dual pairs, their orientation reflecting the twisted boundary condition of Eq.(\ref{mobius}) giving rise to Zamolodchikov periodicity.}
\end{figure}

The evolution of causal patterns can be characterized as a walk of $N-3$ distinguishable particles subject to a set of rules (to be described below) in a triangular region of the past light-cone of the point ${\mathcal A}_{N-1}$. (The reader can visualize the following considerations by looking at the triangular region cut out of the light cone of the point ${\mathcal A}_{5}$ for the $N=6$ case in Figure 14.)
The evolution process starts by putting all of the $N-3$ particles on the right hand side of our triangular region.
This means that all of the particles are lined up on the point curve of the point $N-1$. 
We identify the particle labels $1,2,\dots N-3$ with the label of the other point curve the particle is lying on. 

Now we define a coordinate grid in our triangular region with left/right moving light cone coordinate lines by representing them by red/green segments of the relevant point curves. 
The right moving segments are on the point curves with labels $0,1,\dots N-3$, and the left moving ones are on the ones with labels $2,3,\dots ,N-1$. Let us now relabel the left moving grid lines as: $1,2,\dots N-2$.  
Let us now focus on the red and green segments with identical labels: $1,2,\dots,N-3$.
 The green and red parts of these curves are exchanged after reflecting  from the boundary of the triangular region.
We call these $N-3$ curves the "reflected point curves". Moving one unit {\it down} along the green part will be called a {\it green move}. Similarly moving one unit {\it up} along a red part will be called a {\it red move}.
Notice that since such green and red lines are labelled from $1,2,\dots N-3$ they are naturally {\it ordered}. Then particles executing moves are prioritized according to the magnitude of the curve on which they are located.
For a particle executing red moves of a certain (red) priority an intersecting a green curve will be called {\it occupied} if it is already featuring another particle with a larger (red) priority.

\begin{figure}
{\includegraphics[height=5.5cm]{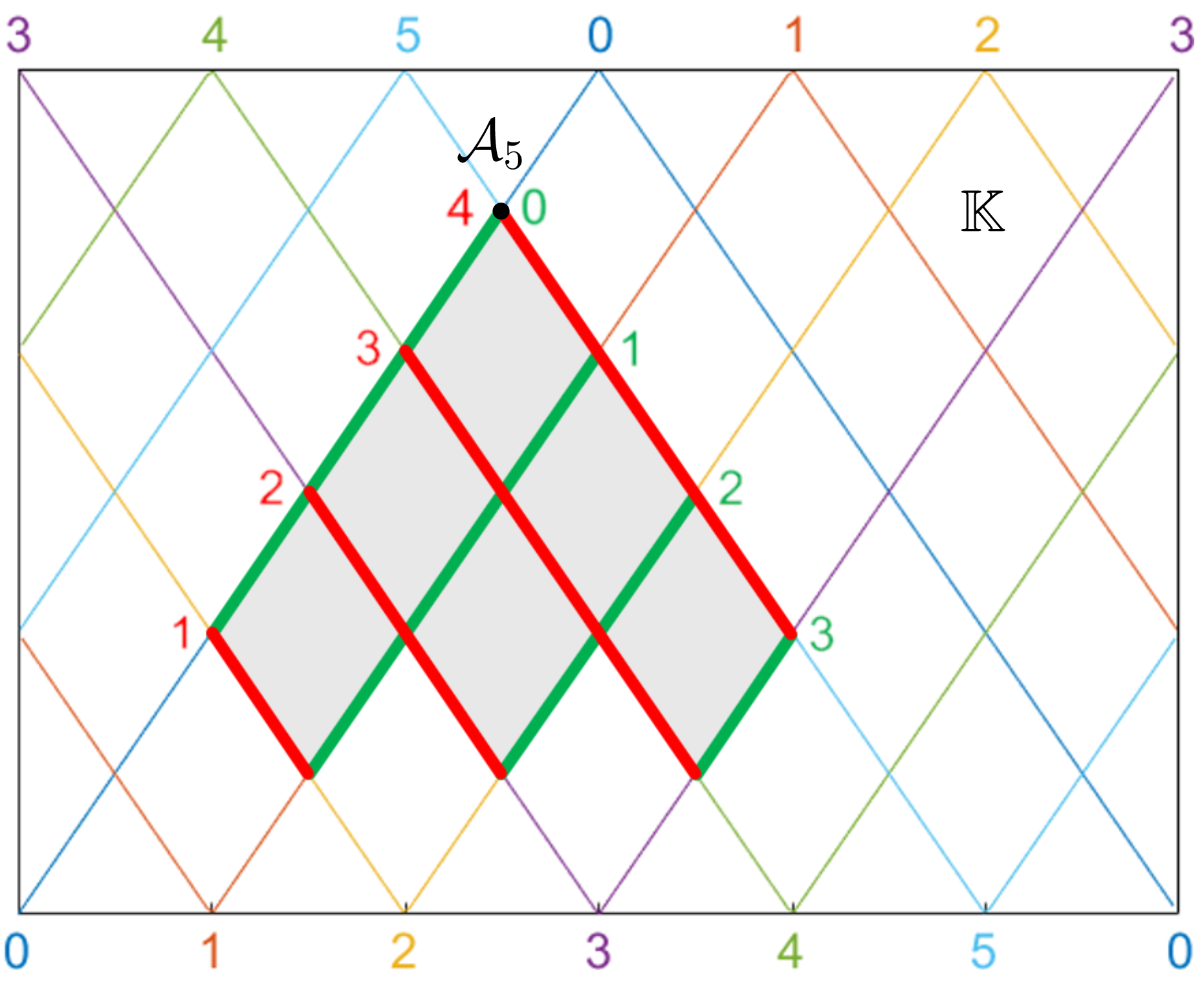}}
\caption{The triangular region of kinematic space where the $N=6$ random walk takes place.}
\end{figure}

Let us call the initial configuration the one when the $N-3$ particles are lined up on the right hand side of our triangular region, and the final one when all of them are lined up on the left hand side of the region. (See the top and bottom configurations of the $N=5$ case of Figure 15.) These configurations correspond to the maximal and minimal elements of the Tamari lattice $T_{N-2}$.
Now the evolution of the initial configuration starts with a random choice of one from the $N-3$ particles. The next step involves applying to this particle a sequence of green moves followed by a sequence of red ones with the particular sequences determined by the following set of rules. {\bf 1}. If the point curve of the particle is not blocked by any other one in the green direction, then the particle walks employing green moves to reach the boundary, then switches there to execute red moves. {\bf 2}. Any walk of a particle executing red moves terminates at the crossing point with an occupied green curve. In the absence of any obstruction for red moves the particle will reach the final (leftmost) position.   
{\bf 3}. A particle executing green moves blocked by another particle switches to executing red moves from the meeting point.
These rules will then result in a new configuration. Now apart from the particle having just finished its walk we have $N-4$ new ones to choose from for the continuation of the evolution.
	After choosing randomly at each step the process terminates at the final configuration. The simplest case of the random walk associated with $T_3$ is illustrated in Figure 15.

It is easy to check that this collection of rules to be applied in $\mathbb K$ corresponds to the usual flip operation of diagonals of geodesic quadrangles in $\mathbb D$.
However, since the location of the $N-3$ particles also encodes the corresponding causal pattern taken together with their area (conditional mutual information) labels, the new representation also encapsulates a visualisation of patterns of entanglement of the CFT vacuum in two dimensional de Sitter space-time.
Another virtue of this geometric approach is that we can immediately read off the components of the $(N-2)$-vectors as some sort of coordinates identifying a particular causal pattern.

In order to show that the $(N-2)$-component vectors are indeed nicely displayed in our triangular regions of $\mathbb K$ where the random walk enfolds first observe that the $(N-2)$th component always equals $N-2$. This correponds to the omnipresent outer bracket in the binary bracketing, or alternatively of our regarding the geodesic ${\bf A}_{N-1}$ as a degenerate diagonal. In $\mathbb K$ this reflects the fact that the point $\mathcal{A}_{N-1}$ being the tip of the light-cone plays a special role in our considerations. Hence one only has to focus on the first $N-3$ components of our vector. 

The next step is to notice that the reflected point curves for the corresponding $N-3$ particles are showing up in a dual role.
Indeed, the green label of these curves can be used to identify the label of the component, and the red label can be used to identify the value in question of that component.
For example for the topmost pattern of Figure 15. all of the particles are having red label equals $3$, and they are on the reflected point curves: number $1$ and $2$. As another example one can consider the leftmost pattern. Here the red labels give $1$ and $3$, and the first one can be associated with point curve number $1$ and the other with number $2$.

One can obtain similar arrangements with three particles for the $N=6$ case. They determine the causal patterns of Figure 13.
For this case the whole collection of $4$-vector labels for $T_4$ can be seen on Figure 13. 
Using these labels the reader can check that the partial order structure of the causal patterns holds.
These labels also fix the positions of the $3$ distinguishable particles in the region of Fig. 14. (Not shown in Figure 13.) 
This produces a figure for $N=6$ similar to the detailed one of Figure 15. of $N=5$.

\begin{figure}
{\includegraphics[width=\columnwidth]{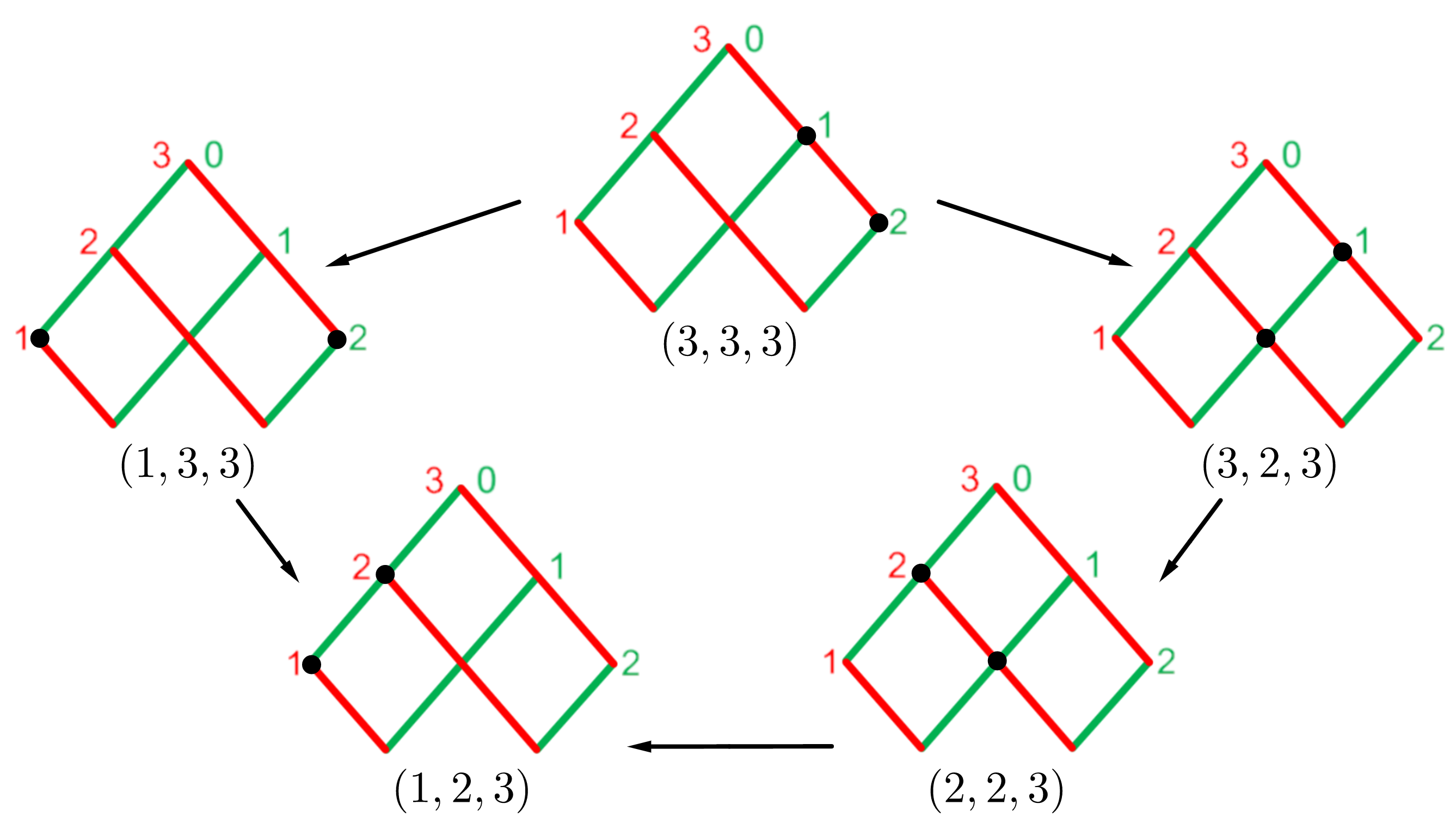}}
\caption{The walk for $N=5$ in kinematic space. Note that in this walk two particles are moving, represented by the two black bullets showing up in these diagrams. In fact there should be a third bullet at the tip of the light cone, which corresponds to a particle never changing its position. However, we omitted this particle from our diagrams. Since the position of this particle, i.e. the position of ${\mathcal A}_4$, is fixed it produces 
at each step of the walk
the same number (namely $3$) for the ($3\equiv 0$ mod $3$) third component of the vector in ${\mathcal V}_3$.}
\end{figure}

Finally we remark that since at the level of such $N-2$-vectors the partial order relation boils down to just componentwise comparison, this result provides a causal-like ordering even for our collection of $N-3$
points representing causal patterns. 
In this respect just recall that for example in Minkowski space-time $x$ causally precedes $y$ if $y-x$ is future-directed null or future-directed timelike. This relation gives the usual partial ordering of space-time and can be written as $x\leq y$. In the same spirit one can {\it define} the causal pattern ${\mathcal T}_1$ causally preceeding the causal pattern ${\mathcal T}_2$ if ${\mathcal T}_1\leq{\mathcal T}_2$. 
Of course then space-like separation arises between two causal patterns when the patterns are {\it not} related with respect to the Tamari order.
Furthermore, thanks to the explicit relationship found  here between causal patterns and conditional mutual informations of boundary regions the Tamari order also introduces a partial order between patterns of entanglement. The representation of collections of  boundary regions as some sort of "space of contexts" via binary bracketings shows that the Tamari ordering provides a natural order on this space.  The elaboration of this interesting idea will be explored in a future work.

\section{Conclusions}
\subsection{Summary}

Using the principle of holography and following the ideas introduced by one of us in a previous paper\cite{Levay} in this work we elaborated on the issue of how patterns of entanglement of the boundary are encoded into the classical geometry of the bulk. In this paper we have choosen the simplest setup namely the ${\rm AdS}_3/{\rm CFT}_2$ correspondence by restricting our attention to the static slice and on patterns of entanglement of the CFT vacuum. Apart from the basic spaces featuring any holographic consideration, namely the bulk and its boundary, we have also made use of kinematic space which is usually regarded as an intermediary between bulk and boundary\cite{Czech1}.
However, apart from translating between the language of quantum information of the boundary $\partial\mathbb D$ to the language of geometry of the bulk $\mathbb D$, the kinematic space $\mathbb K$ is also an interesting space in its own right. Namely, in the static case $\mathbb K$ is just the two dimensional de Sitter space ${\rm dS}_2$, which provides a very simple example of an emergent space-time structure to be understood in quantum entanglement terms. In fact it is a vacuum solution of Einstein's equations hence its connection to the CFT vacuum seems to be of basic importance.

We have shown that using the language provided by the simplest of cluster algebras one can understand ${\rm dS}_2$ as a space of causal patterns encoding patterns of entanglement of the CFT vacuum. The patterns of entanglement are provided by the set of cutoff independent conditional mutual informations associated with binary bracketings of partitions of the boundary into $N$ subregions. The set of all such bracketings defines the space of contexts of $N$ subregions of cardinality $C_{N-2}$.
On the other hand the causal patterns are defined by $C_{N-2}$ configurations of $N-3$ pairwise compatible points in ${\rm dS}_2$.  
Such causal patterns are consisting of $N-3$ causal diamonds with their area labels related to conditional mutual informations and are also connected to the cluster variables. 
Mutations of causal patterns then turn out to be related to the coefficient dynamics\cite{ClusterFZ,Williams} of a cluster algebra of type $A_{N-3}$.

Such mutations consist of elementary flips. At the boundary level they are represented by applying the semi-associative law of right or left shifts of binary bracketings. This corresponds to a flip in the space of contexts. At the bulk level they are represented by flips between the two possible diagonals of a geodesic quadrangle. This corresponds to a flip in the space of triangulations of geodesic $N$-gons. As described in Section VI. at the kinematic space level a flip is represented by an elementary step taken by one of $N-3$ particles executing a random walk on the lattice inside a triangular region provided by the past light cone of a distinguished point.   

As is well-known in the mathematics literature the coefficient dynamics of an $A_n$ cluster algebra is the dynamics of cross ratios. Cross-ratios in turn are related to shear coordinates $t_{j,k}$, which define coordinates for Teichm\"uller space of $\mathbb D$ with $N=n+3$ marked points. Then after a flip in $\mathbb D$ these coordinates are transformed according to the coefficient dynamics of the $A_{n}$ cluster algebra. 
This dynamics is related to a Zamolodchikov $Y$-system of type $(A_n,A_1)$. In Eq. (\ref{iksz}) we have rewritten this dynamics in terms of the $X_{j,k}$ plaquette variables ($j+k\equiv 0$ mod $2$) which are defined in Eq.(\ref{lenyeg}) and related to the usual $Y_{j,k}$ variables and shears via Eq.(\ref{mnemonic}).

We have shown that apart from $X_{j,k}$ with $j+k\equiv 0$ mod $2$ there are dual variables with $j+k\equiv 1$ mod $2$. It turned out that the former set of variables is associated with points ($P$), and the latter with complements of the light cones of such $P$s ($\overline C$) in ${\rm dS}_2$. Moreover, we demonstrated that an interesting relation (Eq.(\ref{huha})) holds between the $P$ and $\overline C$ descriptions. Indeed, the difference between the corresponding mutual informations is related to the geodesic length (proper time) between the time-like separated points in $dS_2$ defining the relevant causal diamond labelled by $P$. This result was based on the fact that the nodes of the $A_n$ Dynkin diagram are of two type: even or odd. Note that since any Dynkin diagram is a tree\cite{Williams} this result generalizes for other type of cluster algebras and cluster dynamics, possibly associated with other boundary quantum states different from the vacuum\cite{Levay}.

For fixed $N$ the space of causal patterns $T_{N-2}$ provides a set of regions covering ${\rm dS}_2$. The collection of such patterns of cardinality $C_{N-2}$ can be identified with the vertices of the associahedron ${\mathcal K}^{N-3}$. The boundaries of the associahedron of different dimensions correspond to different partial triangulations. In particular the collection of facets of cardinality $N(N-3)/2$ can be identified with the collection of Ryu Takayanagi geodesics corresponding to diagonals. 
We have also established on the space of causal patterns $T_{N-2}$ a partial order rendering this space a lattice $(T_{N-2},\leq)$ isomorphic to the Tamari lattice.
This space is also isomorphic to the set ${\mathcal V}_{N-2}$ of $(N-2)$-vectors with their components taken from $\mathbb N$ satisfying the constraints: for $\mu=1,2,\dots ,N-2$ we have $\mu\leq v_{\mu}\leq N-2$, and the one that $\mu \leq\nu\leq v_{\mu}$ implies $v_{\nu}\leq v_{\mu}$.
The partial order in $T_{N-2}$ is then rephrased in ${\mathcal V}_{N-2}$ as componentwise comparison\cite{TamariHuang,Birkhoff}.
We demonstrated that the elements of ${\mathcal V}_{N-2}$ used as coordinates for the elements of $T_{N-2}$ are nicely visualised in the triangular region inside the past light cone of the point ${\mathcal A}_{N-1}$. The first $N-3$ components are identifying $N-3$ points as the locations of the particles subject to the random walk explained in the previous paragraph.

We observed that the Tamari order defines a natural causal ordering in the space of contexts of the boundary.
It is well-known that boundary {\it regions} enjoy a causal order based on the containment relation\cite{Czech1}. Now we have revealed that there is yet another type of causal ordering present in the boundary. It is a partial order which for a fixed $N$ relates different contexts (represented by different binary bracketings) of $N$-fold partitions of the boundary. 
Clearly the context space with this causal order is characterized by the set of mutual conditional informations $I_{j,k}, j=1,2,\dots n+3, k=1,2,\dots n$, and it introduces a new type of holographic entanglement hierarchy taken together with a corresponding evolution governed by the $A_n$ cluster dynamics. The physical meaning of these notions in the holographic context is yet to be explored.

\subsection{Outlook}
Let us finally comment on possible ramifications of our work connecting it to currently explored research topics.
First of all notice that as $N$ goes to infinity, we are getting finer grids on $\mathbb K$ hence the recursion of Eq.(\ref{iksz}) should boil in the continuum limit down to some sort of field equation, with the $X_{j,k}$ variables giving rise to fields $X(\theta,\alpha)$. Due to the (\ref{lenyeg}) relation this approach will produce a field $I(\theta,\alpha)$ for the conditional mutual informations.
What is this field equation? 

It is well-known that the recursion valid for $Y$-systems can be regarded as the discretization of Liouville's equation\cite{Fagyi1,Fagyi2}.
So following the ideas in these papers one is expecting that the relevant field equation is just Liouville equation. However, there are statements\cite{Kashaev} in the literature that though it is easy to see that a formal discretization of Liouville equation indeed yields an Y-system identical to the one of our equation (\ref{zamoan}), however it is not an $Y$-system satisfying Zamolodchikov periodicity.
Since in our setting up Zamolodchikov periodicity is inherently connected to periodicity of kinematic space in order to answer this question some adaptation of the ideas of these papers should be desirable.  The equation in question should be a wave equation in ${\rm dS}_2$ with some source term related to $I(\alpha,\theta)$. 
However, in order to properly implement Zamolodchikov periodicity one also has to take into account the continuum version of Eq.(\ref{mobius}), i.e. a twisted boundary condition. 

Notice that our problem is related to the observation\cite{Myers,BoerMyers} that the one region entanglement entropy is a Liouville field. Then one can conjecture that the entanglement dynamics of the ${\rm CFT}_2$ can be described some sort of $1+1$ dimensional gravity. Though pure Einstein gravity is trivial in $1+1$ dimensions however, Jackiw-Teitelboim (JT) gravity\cite{Jackiw,Teitelboim} is not.  
This idea has already been followed in the literature\cite{Verlinde,Callebaut} with the conclusion that the dynamics in $\mathbb K$ is indeed gravitational and is described by JT-theory. Now one of the equations in JT-theory is just Liouville equation for $S$ of the form\cite{Callebaut}
\begin{equation}
{\partial}_u{\partial}_v\left(\frac{12}{\mathfrak{c}}S\right)=\frac{2}{\delta^2}e^{-\frac{12}{\mathfrak{c}}S}.
\end{equation}
where $\delta$ is an UV-cutoff and $u,v$ is related to $\alpha,\theta$ by Eq.(\ref{uv}).
Hence a naive guess for the corresponding equation to be satisfied by $I(\theta,\alpha)$ would be a cutoff independent analogue of this equation with $S(\theta,\alpha)$ replaced by $I(\theta,\alpha)$. Moreover, one should ensure also that the twisted boundary condition $I(\theta+\pi,\pi-\alpha)=I(\theta,\alpha)$ holds.

There is yet another very interesting connection of our work with the recent research on the scattering amplitudes concerning the bi-adjoint $\phi^3$ theory\cite{Nima1}. According to this theory the tree-level scattering amplitudes are given by the geometry of the associahedron in a space also called by the authors "kinematic space". Moreover, in this seemingly different context quite naturally for the tree level calculations a cluster algebra of type $A_{N-3}$ shows up.
In this approach "kinematic space" is the space of linearly independent Mandelstam  invariants $s_{ij}=2p_i\cdot p_j$ where $i,j=1,2,\dots N$ and the $p_i$ are massless momenta. Since the $s_{ij}$ are not independent it turns out that the dimension of "kinematic space" is just the same as the number of facets of the associahedron ${\mathcal K}^{N-3}$ which is $N(N-3)/2$ the same as the number of Ryu-Takayanagi geodesics corresponding to diagonals in the triangulations of our bulk. Another basis of the same cardinality can be used for the Mandelstam invariants by the planar variables (propagators) $\chi_{i,j}$. Identifying the momenta as $N$ marked points on $\partial\mathbb D$ one realizes that just like the entanglement entropies of boundary regions anchored to the diagonals that are related to the (regularized) geodesic lengths, such planar variables are related to the lengths along the diagonals between $i$ and $j$.
Moreover, from our previous paper\cite{Levay} it is known that for the geodesics comprising the sides of the geodesic $N$-gon by a suitable choice of regulators the entanglement entropies can be scaled away. This can be done thanks to a gauge degree of freedom encoded into the geometry of the space of horocycles and the fact that the associated lambda lengths can be related to the entanglement entropy\cite{Levay}. This corresponds to the fact that the planar variables $\chi_{i,i+1}$ and $\chi_{1,n}$ are vanishing, i.e. they have length zero.   

Now recall Eq.(45) of Ref.\cite{Levay}. According to this equation the entanglement entropies of boundary regions are related to the {\it magnitude} of the lambda lengths of the corresponding geodesics. According to the previous paragraph lambda lengths for {\it diagonal} boundary regions between non adjacent $i$ and $j$ can be related to the planar variables $\chi_{i,j}$. However, unlike entanglement entropies lambda lengths can be negative. Now if we recall the positivity condition Eq.(3.4) of Ref.\cite{Nima1} one realizes that this condition can be interpreted as a constraint coming from physics provided we are willing to switch to a new interpretation of ${\chi}_{ij}$ as a quantity related somehow to entanglement entropies. Now in order to match the dimension of ${\mathcal K}^{N-3}$ and the dimension of the physically relevant subspace of "kinematic space" the authors of Ref.\cite{Nima1} invoke a {\it positivity constraint} of their Eq.(3.6). Now our important observation is that if the planar variables are interpreted as entanglement entropies then this positivity constraint is our strong subadditivity relation of Eq.(\ref{Nimacje}) which is now satisfied automatically! 
Indeed, in this picture the quantities $c_{ij}$ of Ref.\cite{Nima1} are just the conditional mutual informations, the main actors of our paper. Note that strong subadditivity is a highly nontrivial constraint in entanglement theory having somewhat hard to swallow operational meaning\cite{NC} related to entanglement monogamy. However, its geometric meaning is quite transparent in ${\mathbb K}$. (Which is our kinematic space, a $1+1$ dimensional space-time ${\rm dS}_2$, not to be confused with the same terminology used in Ref.\cite{Nima1} for their $N(N-3)/2$ dimensional space where the associahedron lives.)
In Ref.\cite{Levay} we managed to relate boundary entanglement to scattering data in an elementary way, now by working out the fine details of the correspondence just revealed suggests that the analogy between scattering amplitudes and entanglement is probably much deeper.

In this paper we have merely considered the static slice of $AdS_3$. How can we generalize our considerations for a more general setup? Notice in this respect that the basis of our considerations was the link between the conditional mutual informations associated with the boundary and geodesic quadrangles of a static slice of the bulk coming from putting four distinguished points on the boundary.
Clearly when one considers four points of more general locations what one gets is a tetrahedron. Its six edges are geodesics. Now it is known that there are ideal tetrahedra\cite{Gliozzi2} with vertices $a,b,c,d$ labelled by real coordinates with its volume is depending on the cross ratio in the following manner
\begin{equation}
{\rm Vol}(abcd)=L\left(\frac{(a-c)(b-d)}{(b-c)(a-d)}\right)-\frac{\pi^2}{6}
\end{equation}
where $L(x)$ is the Rogers dilogarithm (choosing a sheet in which $\vert{\rm Vol}(abcd)\vert$ having minimal value).
One can then speculate to relate conditional mutual informations of boundary regions to such volumes. Moreover, in this case instead of using triangulations of $\mathbb D$ one have to use triangulations of ${\rm AdS}_3$ by tetrahedra. Then following Refs.\cite{Danciger,Gliozzi2} the analogues of our $X$ and $Y$ variables should be associated to tetrahedra. Moreover, Zamolodchikov's $Y$-systems of type $(A_{N-3},A_1)$ should show up naturally in this treatment\cite{Gliozzi2}.
However, since in this more general case kinematic space is consisting of two copies of ${\rm dS}_2$ spaces\cite{Czech1b,Czech2} at first sight it is not obvious how to generalize our findings.
 
Finally, what about other states of the CFT, different from the vacuum?
As far as this question is concerned we remind the reader that the original context of mathematics where cluster algebras have made their debut was the Teichm\"uller theory of marked Riemann surfaces\cite{Penner,Pennerbook,ClusterFZ,Williams}.
Such surfaces can be uniformized by factorizing $\mathbb D$ by Fuchsian groups $\Gamma$ giving rise to tilings of 
$\mathbb D$ by fundamental domains with their sides identified by certain group elements.
This is the setting where cluster algebras associated with Dynkin diagrams, also others then our one of type $A_{N-3}$, appear. Now $\mathbb D$ can be embedded into ${\rm AdS}_3$ and the action of $\Gamma$ can be extended\cite{Brill,Skenderis} to ${\rm AdS}_3$.
This construction has the well-known physical interpretation of obtaining multiboundary wormhole solutions\cite{Ingemar1,Skenderis} of extremal and non-extremal type generalizing the BTZ black hole\cite{BTZ}.
Now, the holographic dual of such multiboundary wormhole solutions produces an interesting class of CFT states amenable for future study. Such states apparently have entanglement patterns worth studying in kinematic space\cite{Zhang,Zuk}, in a spirit similar to the one of this paper. 
One can then expect that cluster algebras associated with the classical Dynkin diagrams (quivers) appear naturally in this context. Moreover, since such type of cluster algebras are associated with polytopes generalizing the associahedron\cite{Fomin} one can conjecture that in kinematic space (${\rm dS}_2\times {\rm dS}_2$) the new patterns of entanglement are represented in a mathematically natural manner. In particular in the static case these patterns probably show up as certain domains in
${\rm dS}_2$ (as an associated space-time connected to such generalized polytopes) with their random walks adjoined to them.  
The corresponding time evolution in $\mathbb K$ then would dualize to some sort of evolution of entanglement in the boundary CFT.  We hope that the associated order structures might provide some new insight to issues concerning quantum circuits related to holographic codes\cite{Osborne, Yoshida,Czechmera}.

{\it Note added.} While completing this paper we have become aware of a recent paper of Arkani-Hamed et. al. on the topic of scattering amplitudes in the bi-adjoint $\phi^3$ theory\cite{Klassz}, a research direction we commented in our outlook. 
The authors observe that the physical origin for the occurrence of polytopes like the associahedron is associated with causal structures in a space they call "kinematic spacetime" similar to our $\mathbb K$ used here. 
Moreover, they discover the very same structures (cluster algebras, wave equation, causal structures, walk) in their scattering context. Now the challenge is to understand these results in terms of the entanglement picture presented here. 
Hopefully it will also offer a new twist to the idea of holography.

\bigskip
\section{Acknowledgement}

This work was supported by the National Research Development and Innovation Office of Hungary within the Quantum Technology National Excellence Program (Project No. 2017-1.2.1-NKP-2017-0001).

\end{document}